\documentclass[pre,showpacs,twocolumn,nofootinbib]{revtex4}
\usepackage{graphicx}
\usepackage{bm}

\def\zexpo{\theta}
\def\dropletexpo{\omega}
\def\tworG{{\cal G}}
\def\tworQ{{\cal Q}}
\def\tworW{{\cal W}}
\def\tworZ{{\cal Z}}
\def\gatsing{{g_{1}}}
\def\Qsing{\zeta(0)}
\def\avereps{\overline{\varepsilon}}
\def\structures{{\bf S}}
\def\singlezsing{z_0}
\def\tworzsingmolten{z_1}
\def\tworfmolten{f_1}
\def\tworccmolten{s_1}
\def\tworzsingglass{z_2}
\def\tworfglass{f_2}

\def\tworgatsingglass{g_2}
\def\slope{z\prime}
\def\alphsize{K}

\def\Qhat{\widehat{\tworQ}}
\def\qhat{\widehat{q}}
\def\thetahat{\widehat{\theta}}
\def\Pihat{\widehat{\Pi}}
\def\Khat{\widehat{{\cal K}}}

\begin{document}

\title{Statistical mechanics of secondary structures 
formed by random RNA sequences}
\author{R.~Bundschuh}
\altaffiliation[Current address: ]{Department of Physics,
The Ohio State University, 174 W 18th Ave, Columbus, OH 43210-1106, U.S.A.}
\author{T.~Hwa}
\affiliation{
Department of Physics, University of California at San Diego,
9500 Gilman Drive,
La Jolla, CA  92093-0319, U.S.A.}
\date{\today}

\begin{abstract}
The formation of secondary structures by a random RNA sequence
is studied as a model system for the sequence-structure problem 
omnipresent in biopolymers. Several toy energy models are introduced
to allow detailed analytical and numerical studies. First, a two-replica
calculation is performed. By mapping the two-replica problem 
to the denaturation of a single homogeneous RNA in $6$-dimensional 
embedding space,  we show that sequence disorder is perturbatively irrelevant, 
i.e., an RNA molecule with weak sequence disorder is in a {\em molten phase}
where many secondary structures with comparable total energy coexist.
A numerical study of various models at high temperature reproduces behaviors 
characteristic of the molten phase. On the other hand, a scaling argument
based on the extremal statistics of rare regions
can be constructed to show that the low temperature phase is unstable to 
sequence disorder. We performed a detailed numerical study of the 
low temperature phase using the droplet theory as a guide,
and characterized the statistics of large-scale, low-energy 
excitations of the secondary structures from the ground state structure.
We find the excitation energy to grow very slowly (i.e., logarithmically) 
with the length scale of the excitation, suggesting
the existence of a marginal glass phase.
The transition between the low temperature glass phase and the high
temperature molten phase is also characterized numerically. It
is revealed by a change in the coefficient of the logarithmic excitation 
energy, from being disorder dominated to entropy dominated.

\end{abstract}

\maketitle

\section{Introduction}

RNA is an important biopolymer critical to all living systems~\cite{rnagen}
and may be the crucial entity in prebiotic evolution~\cite{rna}.
Like for DNA, there are four types of nucleotides
(or bases) A, C, G, and U which, when polymerized can form double helical
structures consisting of stacks of stable Watson-Crick pairs (A with U
or G with C).  However unlike a long polymer of DNA, which is often
accompanied by a complementary strand and forms otherwise featureless
double helical structures, a polymer of RNA usually ``operates'' in
the single-strand mode. It bends onto itself and forms elaborate
spatial structures in order for bases located on different parts of
the backbone to pair with each other, similar conceptually to 
how the sequence of an amino acid encodes the structure of a protein.

Understanding the encoding of structure from the primary sequence has
been an outstanding problem of theoretical biophysics. Most
theoretical work in the past decade have been focused on the problem
of protein folding, which is very difficult analytically and
numerically~\cite{dill95,woly97,gare97,shak97}.  Here, we study the
problem of RNA folding, specifically the formation of RNA {\em
secondary structures}. For RNA, the restriction to secondary structures
is meaningful due to a separation of energy scales. It is this
restriction that makes the RNA folding problem amenable to detailed
analytical and numerical studies~\cite{higg00}. 
There exist efficient algorithms to compute the
exact partition function of RNA secondary
structures~\cite{zuke81,mcca90,hofa94}. Together with the availability
of carefully measured free energy parameters~\cite{frei86} describing
the formation of various microscopic structures (e.g., stacks, loops,
hairpins, etc.), the probable secondary structures formed by any given
RNA molecule of up to a few thousand bases can be obtained readily.
On the experimental side, RNA molecules of $10^2$ -- $10^5$ bases
in length are available. Furthermore, the restriction to secondary structures 
can be physically enforced
in a salt solution with monovalent ions, e.g., ${\rm Na}^+$,
 so that controlled experiments are in principle possible~\cite{tino99}.  

In this work, we are not concerned with the structure formed by a
specific sequence. Instead, we will study the statistics of secondary
structures formed by the ensemble of {\em long} {\em random} RNA sequences
(of at least a few thousand bases in length in practice).  Such
knowledge may be of value in detecting important structural components
in messenger RNAs which may otherwise be regarded as random from the structural
perspective, in understanding how functional
RNAs arise from random RNA sequences~\cite{rna}, or in
characterizing the response of a long single-stranded DNA molecule to
external pulling forces~\cite{maie00}. More significantly from the
theoretical point of view, the RNA secondary structure problem
presents a rare tractable model of a random heteropolymer where
concrete progress can be made regarding the thermodynamic
properties~\cite{higg00,dege68,higg96,bund99,pagn00,hart01,pagn01}.
Nevertheless, there are many gaps in our understanding.
This paper is a detailed report of our on-going effort  
in this regard.  It provides a self-contained introduction
of the random RNA problem to statistical physicists as a novel
problem of disordered systems, and depicts several approaches we
have tried to characterize this system.

The manuscript is organized as follows: In Sec.~\ref{sec_secondarystructure},
we provide a
detailed introduction to the phenomenology of RNA secondary structure
formation.  We review the key simplifications which form the basis of
efficient computing as well as exact solutions in some cases. We also
review the properties of the ``molten phase", which is the simplest
possible phase of the system assuming sequence disorder is not
relevant. In Sec.~\ref{sec_randomnesshighT}, we consider the effect of
sequence disorder at high temperatures. We show
numerical evidence that the random RNA sequence is in the molten phase
at sufficiently high temperatures, and support this conclusion by
solving the two-replica system which can be regarded as a perturbative
study on the stability of the molten phase. In Sec.~\ref{sec_randomnesslowT},
we provide a scaling
argument, and show why the molten phase should break down at low
enough temperatures. This is followed by a detailed numerical study of
the low temperature regime. We apply the droplet picture and
characterize the statistics of large-scale, low-energy excitations of
the secondary structures from the ground state structure.   Our results
support the existence of a very weak (i.e., marginal) glass phase
characterized by logarithmic excitation energies. Finally, we describe
the intermediate temperature regime where the system makes the transition
from the glass phase to the molten phase. The solution of the two-replica
problem is relegated to the appendices. We present two approaches: In
Appendix~\ref{app_intuitive}, 
we provide a mapping of the two-replica problem to the denaturation 
of an effective single RNA in $6$-dimensional embedding space; this approach 
highlights the connection of the RNA problem to the self-consistent Hartree
theory and should be most natural to field theorists. In
Appendices~\ref{app_solution} and~\ref{app_derivez1},
we present the exact solution. It is 
hoped that the two-replica solution may be helpful in providing the
intuition needed to tackle the full $n$-replica problem.

\section{Review of RNA secondary structure}\label{sec_secondarystructure}

\subsection{Model and definitions}

\subsubsection{Secondary structures}

\begin{figure}[htpb]
\begin{center}
\includegraphics[width=0.6\columnwidth]{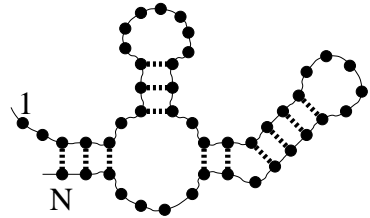}
\caption{Diagramatic representation of an RNA secondary structure: The
solid line symbolizes the backbone of the molecule while the dashed
lines stand for the hydrogen-bonded base pairs formed. The backbone is
shaped such that stems of subsequent base pairs and the loops
connecting or terminating them can be clearly seen. These stems form
double helical structures similar to that of DNA.}
\label{fig_diagrams}
\end{center}
\end{figure}

The secondary structure of an RNA describes the configuration of base
pairings formed by the polymer. If the pairing of the $i^{\rm th}$
and $j^{\rm th}$ bases in a polymer of $N$ total bases is denoted by 
$(i,j)$ with  $1\le i<j\le N$, then each secondary structure
$S$ is defined by a list of such pairings, with each position appearing
at most once in the list, and with the pairs subject to a certain restriction
to be described shortly below. Each such structure can be
represented by a diagram as shown in Fig.~\ref{fig_diagrams}, where
the solid line symbolizes the backbone of the molecule and the dashed
lines stand for base pairings. The structure shown can be divided into
{\em stems} of consecutive base pairs and {\em loops} which connect or
terminate these stems.  In naturally occurring RNA molecules, the stems
typically comprise on the order of five base pairs. They locally form
the same double helical structure as DNA molecules. However, while the
latter typically occur in complementary pairs and bind to each other,
RNA molecules are mostly single-stranded and hence must fold back onto
themselves in order to gain some base pairings.

\begin{figure}[htbp]
\begin{center}
\includegraphics[angle=0,width=1.0\columnwidth]{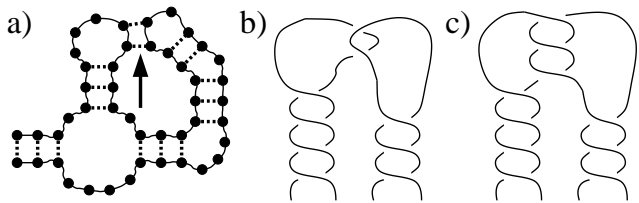}
\caption{Pseudo-knots in  RNA structures: The base pairings
indicated by the arrow in (a) create a pseudo-knot. We 
exclude such configurations in our definition of secondary 
structures: The short pseudo-knots (called
``kissing hairpins'') as shown in (b) 
do not contribute much to the total binding energy, and the long
ones shown in (c) are kinetically forbidden since the
double helical structure would require
threading one end of the molecule through its loops many times.}
\label{fig_pseudoknot}
\end{center}
\end{figure}

\begin{figure}[h]
\begin{center}
\includegraphics[width=0.8\columnwidth]{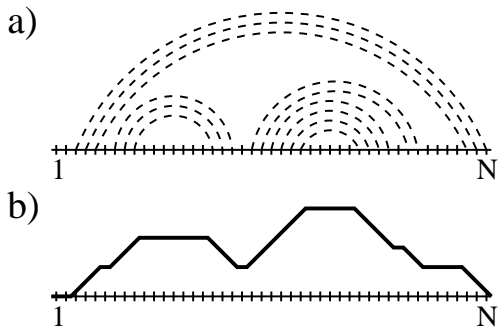}
\caption{Abstract representations of the RNA secondary structure shown
in Fig.~\protect\ref{fig_diagrams}: In
(a) the solid line symbolizes the stretched out backbone of the molecule
while the dashed arches stand for the base pairs formed.  Due to
the no pseudo-knot constraint two arches never cross. (b) shows an equivalent representation
as a ``mountain diagram''. It is a line derived from the arch diagram
by going along the backbone from left to right and going one step up
for every beginning arch, horizontally for each unbound base, and one
step down for each ending arch. Such a mountain always stays above the
baseline and comes back to the baseline at base \protect$N$.}
\label{fig_strreps}
\end{center}
\end{figure}

By a secondary structure, one often considers only the restricted
set of base pairings where any two base pairs
$(i,j)$ and $(k,l)$ in a given secondary structure are either
independent, i.e., $i<j<k<l$, or nested, i.e., $i<k<l<j$. This excludes
the so-called pseudo-knots (as exemplified by Fig.~\ref{fig_pseudoknot})
and makes analytical and numerical studies much more tractable.
For an RNA molecule, the exclusion of pseudo-knots is a reasonable 
approximation because the long pseudo-knots are kinetically
difficult to form, and even the short ones occur infrequently 
in natural RNA structures~\cite{tino99}. The latter is due to their
relatively low binding energies for short sequences 
and the strong electrostatic repulsion of the backbone --- because
the polymer backbone is highly charged and pseudo-knotted configurations
increase the density of the molecule, their formation can be relatively
disfavored in low salt solution. Similarly, the tertiary structures
which involve additional interactions of paired bases are strongly dependent
on electrostatic screening and can be ``turned off" experimentally
by using monovalent salt solution~\cite{tino99}. Indeed, the pseudo-knots 
are often deemed part of the tertiary structure of an RNA molecule.
Throughout this study, we will exclude pseudo-knots in our definition
of secondary structures. Without the pseudo-knots, 
a secondary structure can alternatively be represented by
a diagram of non-crossing arches or by a ``mountain" diagram as shown in
Fig.~\ref{fig_strreps}.

\subsubsection{Interaction energies}\label{sec_intenergies}

In order to calculate Boltzmann factors within an ensemble of
secondary structures, we need to assign an energy
$E[S]$ to each structure $S$. Each secondary structure can be
decomposed into elementary pieces such as the stems of base pairs 
and the connecting loop regions as shown in
Fig.~\ref{fig_diagrams}. A common approach is to 
 assume that the
contributions from these structural elements to the total 
energy are independent of each other and additive.

Within a stem of base pairs, the largest energy contribution is the
{\em stacking energy} between two adjacent base pairs (G-C, A-U, or
G-U), and the total energy of the stem is the sum of stacking energies
over all adjacent base pairs. Since each secondary structure is defined as
a {\em single} state in our ensemble, it is necessary to integrate out
all other microscopic degrees of freedom of the bases within a given 
secondary structure and use an effective energy parameter for each base stacking.
The most convenient one to use is the 
Gibbs free energy of stacking~\cite{frei86}, which contains
an enthalpic term due to base stacking, and an entropic term 
due to the loss of single-stranded degrees of freedom (as well as the
additional conformational change of the backbone and even
 the surrounding water molecules) due to base pairing.
The magnitudes of these stacking free energies actually depend
on the identity of all four bases forming the two base pairs bracketing the
stack and are dependent on temperature themselves. 
While their typical values are on the
order of $k_{\mathrm B}T$ at room temperature, the enthalpic and 
entropic contributions are each on the order
of $10k_{\mathrm B}T$. Thus, upon moderately increasing the
temperature from room temperature to about $80^\circ C$, the stacking
free energies become repulsive and the RNA molecule denatures.

The stacking free energies account for most but not all of the entropic 
terms for a given secondary structure. There is an additional (logarithmic)
``loop energy'' term associated with the 
entropy loss of each {\em closed loop} of single-stranded RNA formed by
the secondary structure, as well as the energy necessary to bend the single
strand. All of these energy parameters have been measured  in
great detail~\cite{frei86}. When incorporated into an efficient dynamic
programming algorithm (to be described below), they can rather
successfully {\em predict} the secondary
structures of many  RNA molecules of up to several hundred bases in 
length~\cite{zuke81,mcca90,hofa94}.

In this paper, we investigate the statistical properties of long,
random RNA sequences far below the denaturation temperature. We are
interested in generic issues such as the existence of a glass phase
and various scaling properties. Guided by experiences with other
disordered systems~\cite{youn98}, we believe these generic properties 
of the system should not depend on the specific choice of the model
details. Since the full model used in
Refs~\cite{zuke81,mcca90,hofa94} makes analytical and numerical
studies unnecessarily clumsy, we will examine a number of simplified 
models, while preserving the most essential feature of the system, namely, the
pattern of matches and mismatches between different positions of the
sequence.

As in the realistic model described above, we choose our
reference energy to be the unbound state, so that each unbound base in
a secondary structure is assigned the energy $0$. We will neglect the
logarithmic loop energy, which is important very close to the
denaturation transition~\cite{moro01} where the average binding energy
is close to zero, but not far below the denaturation temperature where
most bases are paired.  Moreover, we will radically simplify the energy
rules for base pairing: We neglect the stacking energies and instead
associate an interaction energy $\varepsilon_{i,j}$ with every pairing
$(i,j)$. Thus,
\begin{equation}\label{eq_energy}
E[S]=\sum_{(i,j)\in S}\varepsilon_{i,j}
\end{equation}
is the total energy of the structure $S$. 

Within this model, it remains to be decided how to choose the energy 
parameters $\varepsilon_{i,j}$'s. One possibility is to choose each 
of the bases $b_1\ldots b_N$ randomly from the `alphabet' set $\{A,
C,G,U\}$ and then assign
\begin{equation}\label{eq_energychoice}
\varepsilon_{i,j}=\left\{\begin{array}{rl}-u_{\mathrm{m}}&
\mbox{$b_i$--$b_j$ is a Watson-Crick base pair}\\
u_{\mathrm{mm}}&\mbox{otherwise}\end{array}\right.
\end{equation}
with $u_{\mathrm{m}}, u_{\mathrm{mm}}>0$ being the match or mismatch energy
respectively. Here, the value of $u_{\mathrm{mm}}$ is actually not essential as long
as it is repulsive, since the two bases always have the energetically
preferred option to not bind at all.
Thus the energetics of the system is set by $u_{\mathrm{m}}$.
In our numerical study to be reported in Secs.~\ref{sec_randomnesshighT}
and~\ref{sec_randomnesslowT}, we will primarily
use this model\footnote{Note that as a toy model, there is no reason why 
the alphabet size of the bases needs to be $4$ (as long as it is larger than 
$2$ as explained below). Indeed the alphabet size and the choice of the
matching rule can be used as tuning parameters to change the strength of 
sequence disorder. But in our study, we choose to minimize the number of 
parameters and tune the effective strength of disorder by changing 
temperature.} with $u_{\mathrm{m}} = u_{\mathrm{mm}} =1$.
We will refer to this as the ``sequence disorder'' model.

For analytical calculations, it is preferable to treat all the
$\varepsilon_{i,j}$'s as {\em independent} identically distributed
random variables, i.e., to assume
\begin{equation}\label{eq_uncorrelated}
\rho[\{\varepsilon_{i,j}\}]=\prod_{1\le i<j\le N}\rho(\varepsilon_{i,j})
\end{equation}
for the joint distribution function $\rho[\{\varepsilon_{i,j}\}]$ of all
the $\varepsilon_{i,j}$'s. This choice neglects the correlations
between $\varepsilon_{i,j}$ and $\varepsilon_{i,k}$ which are
generated through the shared base $b_i$; it is an additional
approximation on the model (\ref{eq_energychoice}). However, we do not
anticipate universal quantities to depend on such subtle correlation
of the $\varepsilon_{i,j}$'s. This  will be tested numerically
by comparing the behavior of the model (\ref{eq_energychoice}) with
that of the model defined by Eq.~(\ref{eq_uncorrelated}) together with
\begin{equation}\label{eq_uncorrelateddisorder}
\rho(\varepsilon) = \frac14\delta(\varepsilon+u_{\mathrm{m}})
+ \frac34\delta(\varepsilon-u_{\mathrm{mm}}).
\end{equation}
This distribution is chosen to mimic the random sequence model 
(\ref{eq_energychoice}) with a 4-letter alphabet, but it does not
contain any correlation between the different $\varepsilon_{i,j}$'s.
We will refer to this model as defined by Eqs.~(\ref{eq_uncorrelated}) 
and (\ref{eq_uncorrelateddisorder}) as the ``energy disorder'' model.

In the actual analytical calculations, we will go even one step further
and take the $\varepsilon_{i,j}$ to be {\em Gaussian} random
variables specified by
\begin{equation}\label{eq_gaussiandisorder}
\rho(\varepsilon) = \frac{1}{\sqrt{2\pi D}}\, e^{-(\varepsilon
-\avereps)^2/2D}
\end{equation}
where $\avereps$ is the average binding energy and $D$ is the variance.
In this model (referred to below as ``Gaussian disorder" model,)
the parameter $D$ provides us with a
convenient measure of the disorder strength.  Again,
universal quantities should not depend on the choice of the 
distribution functions. 
We will test this directly by performing numerical studies
for these Gaussian random energies, with
\begin{equation}
\avereps=-\frac14u_{\mathrm{m}}+\frac34u_{\mathrm{mm}}\quad\mbox{and}\quad
D=\frac{3}{16}(u_{\mathrm{m}}+u_{\mathrm{mm}})^2
\end{equation}
chosen to match the first two moments of the distribution
Eq.~(\ref{eq_uncorrelateddisorder}).

In contrast to prior numerical studies~\cite{pagn00}, we do not
exclude base pairing between neighboring bases $(i,i+1)$, i.e., 
we do not set a minimal allowed length for the hairpins\footnote{We did
however repeat most of the numerical studies presented in this paper
with a minimal hairpin size of $1$. Since the results are qualitatively
identical to the results of the simpler model presented here, we do
not show this data.}. Setting 
a constraint on the minimal hairpin length would make the analytical
study much more cumbersome. However, in the study
by Pagnani \textit{et al.}~\cite{pagn00}, it has been argued that the
system will not be frustrated (and hence will not form a glass) 
without this additional constraint. We believe this is an artifact of the 
2-letter alphabet used by Pagnani \textit{et al.} in order to generate the 
binding energy $\varepsilon_{i,j}$'s via a rule similar to 
Eq.~(\ref{eq_energychoice}): It is simple to see that for any 2-letter
sequence in which the like letters repel and unlike letters attract, 
one can always find the minimal total binding energy by 
pairing up neighboring bases of opposite
types and removing them from the sequence if no additional
constraints such as the minimal hairpin length are enforced.
As we will discuss in detail in Sec.~\ref{sec_proofglass} this is not
a problem if the alphabet size is larger than 2.
Thus, in our study, we use the sequence disorder
model with a 4-letter alphabet, or the energy disorder model, without
enforcing the minimal hairpin length constraint. While the minimal hairpin
length (of 3 bases) is known for real RNA folding, it should not change
the universal properties of long RNA sequences.

\subsubsection{Partition function}

Once the energy of each secondary structure is defined, we can
study the partition function
\begin{equation}
Z(N)=\sum_{S\in\structures(N)}e^{-\beta E[S]}
\end{equation}
of the molecule where $\structures(N)$ denotes the set of {\em all} allowed
secondary structures of a polymer of $N$ bases, and $\beta=1/k_BT$. 
To calculate this
partition function, it is useful to study the {\em restricted}
partition function $Z_{i,j}$ of the substrand from position $i$ to position
$j$ of the RNA molecule.  Given the model (\ref{eq_energy}), the
restricted partition functions can be split up according to the
possible pairings of position $j$. This leads to the recursive
equation~\cite{higg96,dege68,wate78}
\begin{equation}
Z_{i,j}=Z_{i,j-1} +
\sum_{k=i}^{j-1} Z_{i,k-1}\cdot e^{-\beta\varepsilon_{k,j}} \cdot
Z_{k+1,j-1}\label{eq_partfuncrec}
\end{equation}
with $Z(N)=Z_{1,N}$ being the total partition function of the
molecule. In terms of the arch diagrams introduced in
Fig.~\ref{fig_strreps}(a) this can be represented as
\begin{equation}
\parbox{0.9\columnwidth}{\includegraphics[width=0.9\columnwidth]{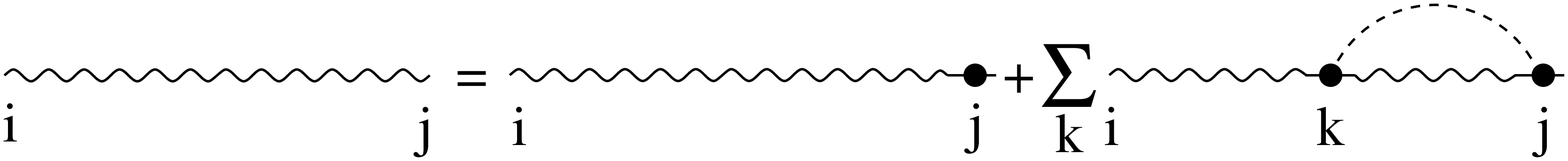}}
\end{equation}
where the wavy lines stand for the restricted partition functions.
This is easily recognized as a Hartree equation.  Since the restricted
partition functions on the right hand side of this equation all
correspond to shorter pieces of the RNA molecule than the left hand
side, this equation allows one to calculate the exact partition
function of an RNA molecule of length $N$ with {\em arbitrary}
interactions $\varepsilon_{i,j}$ in $O(N^3)$ time. This is
accomplished by starting with the partition functions for single bases
and recursively applying Eq.~(\ref{eq_partfuncrec}), and is known as a
dynamic programming algorithm~\cite{mcca90,wate78}.  This algorithm
allows one to compute numerically the partition function involving all
secondary structures, for arbitrary RNA molecules of up to $N\approx
10,000$ bases. It also forms the basis of analytical approaches to the
problem as we will see shortly.

\subsubsection{Physical observables}\label{sec_observables}

Apart from the partition function itself, we will use additional
observables in order to characterize the behavior of RNA secondary
structures. One such quantity of interest is the
binding probability $P_{i,j}$, i.e., the probability that positions $i$
and $j$ are paired given the $\varepsilon_{i,j}$'s,
\begin{equation}\label{eq_returnprobability}
P_{i,j}%\equiv\frac{\widetilde{Z}_{i,j}}{Z_{1,N}}
\equiv\frac{e^{-\beta\varepsilon_{i,j}}Z_{i+1,j-1}Z_{j+1,i-1}}{Z_{1,N}}
\end{equation}
where $Z_{i+1,j-1}$ is given by the recursion
equation~(\ref{eq_partfuncrec}) and $Z_{j+1,i-1}$ is the partition
function of the sequence $b_{j+1}b_{j+2}\ldots b_Nb_1\ldots b_{i-2}b_{i-1}$.
The latter can be calculated as the quantity $Z_{j+1,N+i-1}$ when applying
the recursion Eq.~(\ref{eq_partfuncrec}) to the duplicated sequence
$b_1\ldots b_Nb_1\ldots b_N$. Thus, all $N(N-1)/2$ such constraint
partition functions can be calculated with the same recursion in
$O(N^3)$ time. The logarithms
\begin{equation}\label{eq_pinchfreeenergy}
\Delta F_{i,j}=-k_{\mathrm{B}}T\ln P_{i,j}
\end{equation}
of these binding probabilities have a natural interpretation: they can
be read as the ``pinching free energies'', i.e., as the free energy
cost of a pinch between positions $i$ and $j$ and the unperturbed
state. We will make extensive use of this concept of pinched
structures in our discussion of the low temperature behavior of RNA
secondary structures in Sec.~\ref{sec_randomnesslowT}.  In our numerical
investigations, we will choose as a representative of all the pinching
energies for different positions by
\begin{equation}\label{eq_largestpinchfreeenergy}
\Delta F(N)\equiv\Delta F_{1,N/2+1}
\end{equation}
which is the free energy cost
of the largest possible pinch that splits the molecule of length $N$
into two pieces of length $N/2-1$ each.

\begin{figure}[hbtp]
\begin{center}
\includegraphics[angle=0,width=1.0\columnwidth]{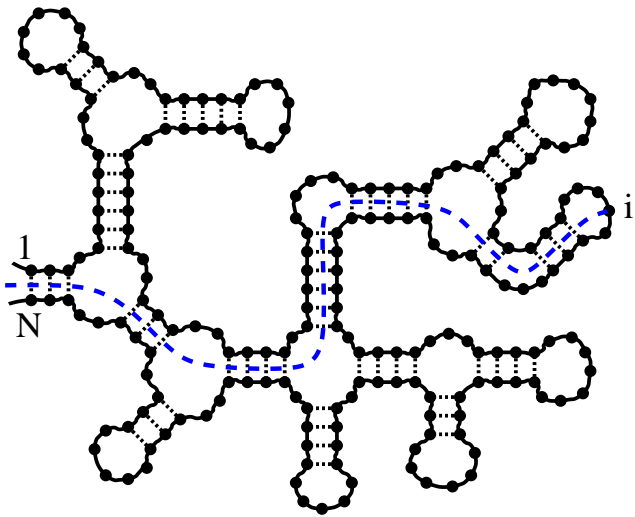}
\caption{Definition of the ``size profile'' \protect$h_i$ of a
secondary RNA structure: The size profile measures the
extension of the structure if drawn as a planar diagram. As an intrinsic
definition of \protect$h_i$ which captures this notion of the size of a
secondary structure at position \protect$i$, we use the ``ladder
distance'' of base \protect$i$ from the end of the molecule, i.e.,
the number of base pairs which have to be crossed when connecting position
\protect$i$ to
position \protect$1$ along the folded structure as indicated
by the dashed line.}
\label{fig_hdef}
\end{center}
\end{figure}

Another quantity which describes a secondary structures is its ``size
profile''. As an intrinsic measure of the size of a given secondary
structure $S$, we use the ``ladder distance'' $h_i(S)$ between the base at
position $1$ and the base at position $i$, which is the the number of
pairings (or ladders) one has to cross to go from a pair involving base
$1$ to the base $i$; see Fig.~\ref{fig_hdef}. It can be 
defined for each secondary structure $S$ as the total number of pairings
$(k,k')\in S$ that bracket $i$, i.e.,
\begin{equation}\label{eq_defmountainheight}
h_i(S)\equiv|\{(k,k')\in S|k<i\le k'\}|.
\end{equation}
This quantity can be very easily visualized as
the ``height'' at position $i$ of
the mountain representation of the secondary structure $S$ as shown in
Fig.~\ref{fig_strreps}(b). A quantity characterizing the full ensemble
of secondary structures is the {\em thermal average} $\langle
h_i\rangle$ of this size profile over all secondary structures with
their respective Boltzmann factors; it can be straightforwardly
calculated from the probabilities $P_{k,k'}$ as
\begin{equation}
\langle h_i\rangle = \sum_{k=1}^{i-1}\sum_{k'=i}^N P_{k,k'}.
\end{equation}
Since we expect all positions in the sequence to behave in a similar
way, in our numerics we will summarize the properties of the
size profile by the ladder distance from the first to the middle
base, i.e., we will study
\begin{equation}\label{eq_hdef}
\langle h\rangle\equiv\langle h_{N/2+1}\rangle
\end{equation}
as a quantity representing the overall ``size'' of an ensemble
of secondary structures.

\subsection{The molten phase}\label{sec_molten}

\subsubsection{Definition of the molten phase}

If sequence disorder does not play an important role, 
we may describe the RNA
molecule by replacing all the binding energies $\varepsilon_{i,j}$ by
some effective value $\varepsilon_0<0$.  As we will see later, this
will be an adequate description of our random RNA models at high enough
temperatures (but before denaturation.) For the real RNAs, this provides
a coarse grained description of repetitive, self-complementary sequences, 
e.g., CAGCAG...CAG, which are involved in a number of
diseases~\cite{mita97}. We will refer to RNA which is
well described by this model without sequence disorder as being in the
``molten'' phase. It serves as a starting point for modeling non-specific
self-binding of RNA molecules, and its
properties will form the basis of our study of the random RNA at low
temperatures.  
\subsubsection{Partition function}\label{sec_moltenpartfunc}

Since in the absence of sequence disorder, the energy of a structure
$S$ depends only on the number of paired bases $|S|$ of this structure,
we can write the partition function in the molten phase as
\begin{equation}
Z(N)=\sum_{S\in\structures(N)}\exp\left[-\beta\varepsilon_0|S|\right]
\end{equation}
The partition functions of the sub-strands $Z_{i,j}$ 
become  translationally invariant and can be written as
\begin{equation}
Z_{i,j}= G(j-i+2)\label{eq_G0}
\end{equation}
where $G(N)$ is only a function of the length $N$.
The recursion equation~(\ref{eq_partfuncrec}) then takes the form
\begin{equation}\label{eq_g0recursion}
G(N+1)=G(N) + q \sum_{k=1}^{N-1} G(k)\cdot G(N-k),
\end{equation}
where
\begin{equation}
q\equiv e^{-\beta\varepsilon_0}. 
\end{equation}
Upon introducing the $z$-transform
\begin{equation}
\widehat{G}(z)=\sum_{N=1}^\infty G(N) z^{-N},
\end{equation}
the convolution can be eliminated and the recursion equation turns
into a quadratic equation
\begin{equation}
z\widehat{G}(z)-1=\widehat{G}(z)+q\widehat{G}^2(z)
\end{equation}
with the solution
\begin{equation}\label{eq_singlez}
\widehat{G}(z)=\frac{z-1-\sqrt{(z-1)^2-4q}}{2q}.
\end{equation}
Performing the inverse $z$-transformation in the saddle point approximation
yields the expression~\cite{dege68,bund99,wate78}
\begin{equation}\label{eq_moltenz}
G(N)\approx A_0(q) N^{-\zexpo_0} \singlezsing^N(q)
\end{equation}
in the limit of large $N$,
with the exponent $\zexpo_0=3/2$ and the non-universal quantities
$\singlezsing(q)=1+2\sqrt{q}$ and $A_0(q)=[(1+2\sqrt{q})/4\pi
q^{3/2}]^{1/2}$.

This result characterizes the state of the RNA where a large number of
different secondary structures of equal energy coexist in the
thermodynamic ensemble, and the partition function is completely
dominated by the configurational entropy of these secondary
structures.  While the result is derived specifically for the special
case $\varepsilon_{i,j} =\varepsilon_0$, we will argue below that it is
applicable also to random $\varepsilon_{i,j}$'s at sufficiently high
temperatures, in the sense that for long RNA molecules, the partition
function is dominated by an exponentially large number of secondary
structures all having {\em comparable} energies (within $O(k_BT)$)
that are smoothly related to each other in configuration space.
The latter is what we meant by the ``molten phase''.

\subsubsection{Scaling behavior}\label{sec_scaling}

The exponent $\zexpo_0=3/2$ is an example of a scaling exponent
characteristic of the molten phase. This and other exponents can be
derived in a geometric way by the ``mountain'' representation of
secondary structures as illustrated in Fig.~\ref{fig_strreps}(b). Each
such mountain corresponds to exactly one secondary structure. In the
molten phase, the weight of a secondary structure $S$ is simply given
by $q^{|S|}$. This can be represented in the mountain picture by
assigning a weight of $q^{1/2}$ to every upward and downward step and
a weight of $1$ to every horizontal step.  Since the only constraints
on these mountains are (i) staying above the baseline, and (ii)
returning to the baseline at the end, the partition function of an RNA
of length $N$ is then simply that of a random walk of $N$ steps,
constrained to start from and return to the origin, in the presence of
a {\em hard wall} at the origin, with the above weights ($\sqrt{q}$ or
$1$) assigned to each allowed step.  This partition function is
well-known to have the characteristic $N^{-3/2}$ behavior which we
formally derived in the last section~\cite{fell50}.

In this framework, it also becomes obvious why imposing a minimal
hairpin length does not change the universal behavior of RNA at least
in this molten phase: If the minimal allowed size of a hairpin is $s$,
this enforces a potentially strong {\em penalty} for the formation of
a hairpin, since with every hairpin $s$ bases are denied the
possibility of gaining energy by base pairing. This tends to make
branchings less favorable and thus leads to longer stems. However,
this additional constraint translates in the mountain representation
into the rule that an upwards step may not be followed by a downwards
step within the next $s$ steps. This is clearly a {\em local}
modification of the random walk. Thus, it does not change universal
quantities although the above mentioned suppression of branchings will
require much longer sequences in order to observe the asymptotic
universal behavior. For real RNA parameters, the crossover length
is very long due to this effect.
For example, it is several hundred nucleotides for 
the CAG repeat, and even longer for some other repeats.

Another characteristic exponent describes the scaling of the ladder size
$\langle h\rangle$ with the sequence length $N$.
As already mentioned in its definition (\ref{eq_hdef}),
$\langle h\rangle$ is equivalent to the average 
``height'' of the mid-point of the sequence in the mountain picture.
In the molten phase, the random walk analogy immediately yields the result
\begin{equation}\label{eq_hscaling}
\langle h\rangle_0 \sim N^{1/2},
\end{equation}
where $\langle ... \rangle_0$ denotes ensemble
average in the molten phase.

As should be clear from the
coarse-grained view depicted in Fig.~\ref{fig_hdef}, the ensemble of RNA
secondary structures in the molten phase can be mapped directly to the 
ensemble of {\em branched polymers}. These branched polymers are 
{\em rooted} at the bases $i=1$ and $i=N$ of the RNA. 
In this context,  $\zexpo_0=3/2$ is known as the
configuration exponent of the rooted branched
polymer~\cite{lube81}. Additionally from the result (\ref{eq_hscaling}), 
we see that the ladder length of the branched polymer scales
\footnote{For a real branched polymer, each branch will have a
spatial extension
which scales as the square root of its ladder length (in the absence
of excluded volume interaction). Then the typical spatial extension 
of a branched polymer  scales as $N^{1/4}$, a well-known result for 
the branched polymer in the absence of self-avoidance~\cite{lube81}.}
as $N^{1/2}$.
Because of the very visual analogy of the secondary structures to
 branched polymer, we refer to the configurational entropy of 
the secondary structures as the ``branching entropy''.

Finally, the binding probabilities $P_{i,j}$ defined in
Eq.~(\ref{eq_returnprobability})
only depend on the distance $|i-j|$ in the molten phase, i.e., 
$P_{i,j} = p(|i-j|)$. The behavior of this function 
can be derived explicitly by inserting the result
Eq.~(\ref{eq_moltenz}) for the partition function into
Eq.~(\ref{eq_returnprobability}). Alternatively, one just needs to recognize
that $p(\ell)$ corresponds in the random walk analogy
to the first-return probability of a random walk after $\ell$-steps.
In either case, one finds the result
\begin{equation}\label{eq_moltenreturn}
p(\ell)\sim \frac{\ell^{-\frac32}(N-\ell)^{-\frac32}}{N^{-\frac32}},
\end{equation}
i.e., the return probability decays with the separation $\ell$ of the
two bases as a power law with the configuration exponent
$\zexpo_0=3/2$. For the pinching free energy $\Delta F(N)$, we simply
set $\ell=N/2$ and obtain
\begin{equation}\label{eq_moltenpinch}
\Delta F_0=\frac{3}{2}k_{\mathrm{B}}T\ln N
\end{equation}
for large $N$, i.e., it scales {\em logarithmically} in the molten phase.
 This logarithmic
dependence merely reflects the loss in branching entropy due to the
pinching constraint and is a manifestation of the configuration 
exponent $\zexpo_0=3/2$.

\section{Effect of Sequence Randomness: High Temperature
Behavior}\label{sec_randomnesshighT}

There are in principle three
different scenaria for the behavior of long random RNA sequences. (i) Disorder
is irrelevant at any finite temperature, so that  the molten
phase description presented in Sec.~\ref{sec_molten} applies to
long RNAs at all temperatures. (ii) Disorder is relevant at all temperatures,
and the molten phase description 
would be completely inadequate. (iii) There is a finite temperature
$T_{\mathrm{g}}$ above which the molten description of random RNA is
correct, while below  $T_{\mathrm{g}}$ a qualitatively different description
is needed. In accordance with the statistical physics literature, 
we will refer to the non-molten phase as the glass phase, and 
$T_{\mathrm{g}}$ as the glass transition. The purpose of the study is to 
determine which of these three scenaria is actually
realized, and to characterize the glass phase 
if either (ii) or (iii) is realized.

In this section, we study the high temperature behavior and
demonstrate that the molten phase is stable with
respect to weak sequence disorder. This ensures that the molten description of
RNA given in Sec.~\ref{sec_molten} is at least valid at high enough
temperatures, thereby ruling out scenario (ii). 
We will address the question of whether there is a glass phase
at low but finite temperatures in Sec.~\ref{sec_randomnesslowT}.

\subsection{Numerics}\label{sec_hightnumerics}

Before we engage in detailed calculations, we want to convince
ourselves with the help of some numerics that weak disorder does not
destroy the molten phase. To this end, we study the observables
introduced in Sec.~\ref{sec_observables}. We generate a large number
of disorder configurations, i.e., interaction energies
$\varepsilon_{i,j}$ using the 3 models introduced in
Sec.~\ref{sec_intenergies}: sequence disorder,
energy disorder, and Gaussian disorder as described by
Eq.~(\ref{eq_energychoice}),~Eqs.~(\ref{eq_uncorrelated}) 
and (\ref{eq_uncorrelateddisorder}),
and Eqs.~(\ref{eq_uncorrelated}) and (\ref{eq_gaussiandisorder}) respectively, 
with $u_{\mathrm{m}}=u_{\mathrm{mm}}=1$. Then, we calculate the observables
$\langle h\rangle$ and $\Delta F(N)$ for each disorder configuration at
the relatively large temperature of $k_BT=2$ and
average the obtained values over many disorder configurations. In
order to keep the numerical effort manageable, we average over
$10,000$ random sequences for $N\in\{10,20,40,80,160,320\}$, over
$2,000$ sequences for $N=640$, and over $1,000$ sequences for
$N=1,280$ and $N=2,560$.

\begin{figure}[htpb]
\begin{center}
\includegraphics[width=1.0\columnwidth]{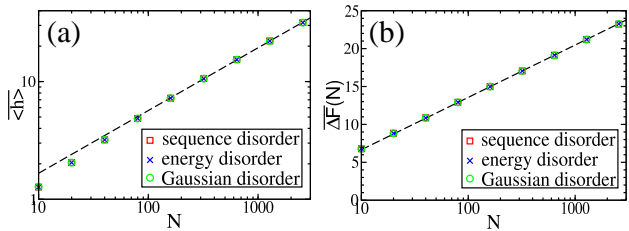}
\caption{Scaling in the molten phase: These two plots show the
dependence of several characteristic quantities of RNA secondary
structures on the length \protect$N$ of the sequence at
\protect$k_BT=2$. Each plot shows data for three different choices of
disorder according to Eqs.~(\protect\ref{eq_energychoice}),
(\protect\ref{eq_uncorrelateddisorder}),
and~(\protect\ref{eq_gaussiandisorder}). (a) shows the scaling of the average
size \protect$\overline{\langle h\rangle}$ and the dashed line is the best fit
\protect$\overline{\langle h\rangle}\sim N^{0.54}$ to a power law.
(b) shows the free energy of the largest pinch as defined
in~(\protect\ref{eq_largestpinchfreeenergy}). The dashed line is up to an
additive constant the logarithmic behavior \protect$\frac32\times2\ln
N$ predicted in Eq.~(\protect\ref{eq_moltenpinch}). The statistical 
fluctuations are smaller than the size of the symbols  in both plots. 
All plots suggest that the behavior of RNA secondary
structures at high temperatures is well described by the molten phase
picture and independent of the disorder.}
\label{fig_moltennumerics}
\end{center}
\end{figure}

Fig.~\ref{fig_moltennumerics} shows the results; disorder averaged
quantities are denoted by an overline throughout the text. We see that the data for
$\overline{\langle h\rangle}$ follows a power law with a fitted
exponent $\overline{\langle h\rangle}\sim N^{0.54}$, with the exponent value
decreasing for larger $N$'s. This result is consistent with the
prediction Eq.~(\ref{eq_hscaling}) for the molten phase. Also, the
pinching free energy follows the predicted logarithmic behavior
Eq.~(\ref{eq_moltenpinch})
without any noticeable difference between the three choices of disorder.
Taken together, these results indicate that the three models of disorder 
belong to the same universality class, i.e., the molten phase description
of the uniformly attracting RNA, at high temperatures.

\subsection{The replica calculation}\label{sec_replica}

Now, we will establish the stability of the molten phase against weak
disorder by an analytical argument. We will use Gaussian disorder
characterized by Eqs.~(\ref{eq_uncorrelated})
and~(\ref{eq_gaussiandisorder}).  As we have shown above, the different
microscopic models of binding energies all yield the same scaling
behaviors. With the uncorrelated Gaussian energies, it is possible to
perform the ensemble average of the partition function
$\overline{Z^n}$ of $n$ RNA molecules sharing the same disorder. The
disorder-averaged free energy can then in principle be obtained via
the ``replica-trick'' $\overline{\ln Z} = \lim_{n\to 0}
(\overline{Z^n}-1)/n$, by solving the $n$-replica problem~\cite{edwa75}.

The $n$-replica partition function can be written down formally as
\begin{widetext}
\begin{eqnarray*}
\overline{Z^n}&=&\sum_{\{S_1\}}\ldots\sum_{\{S_n\}}
\overline{\exp\left[-\sum_{k=1}^n\sum_{(i,j)\in S_k}\beta\varepsilon_{i,j}
\right]}\\
&=&\sum_{\{S_1\}}\ldots\sum_{\{S_n\}}
\prod_{k=1}^n\exp\left[-\beta\avereps|S_k|\right]
\exp\left[\frac{1}{2}\beta^2\sum_{k=1}^n\sum_{l=1}^n\sum_{(i,j)\in S_k}
\sum_{(r,s)\in S_l}
\overline{(\varepsilon_{i,j}-\avereps)(\varepsilon_{r,s}-\avereps)}
\right]\\
&=&\sum_{\{S_1\}}\ldots\sum_{\{S_n\}}\prod_{k=1}^n
\exp\left[-\beta\avereps|S_k|\right]
\exp\left[\frac{1}{2}\beta^2D\sum_{k=1}^n\sum_{l=1}^n|S_k\cap S_l|\right]\\
&=&\sum_{\{S_1\}}\ldots\sum_{\{S_n\}}\prod_{k=1}^n
q^{|S_k|} \prod_{1\le k<l\le n}\widetilde{q}^{|S_k\cap S_l|}
%W_{\beta(\varepsilon_0-\beta D/2)}(S_k)
%\exp\left[\beta^2D\sum_{1\le k<l\le n}|S_k\cap S_l|\right],
\end{eqnarray*}
\end{widetext}
where 
\begin{equation}
q\equiv\exp\left(-\beta\avereps+\frac{1}{2}\beta^2D\right)
\quad\mbox{and}\quad\widetilde{q}\equiv\exp\left(\beta^2D\right)
\end{equation}
are the two relevant ``Boltzmann factors''.  This effective partition
function has a simple physical interpretation: It describes $n$ RNA
molecules subject to a {\em homogeneous} attraction with effective
interaction energy $\varepsilon_0 = \avereps-\frac{1}{2}\beta D$
between any two bases of the same molecule. As before, this effective
attraction is characterized by the factor $q$. In addition, there is
an inter-replica attraction characterized by the factor
$\widetilde{q}$ for each bond {\em shared} between any pair of
replicas. The inter-replica attraction is induced by the same sequence
disorder shared by all replicas. For example, if the base composition
in one piece of the strand matches particularly well with another
piece, then there is a tendency to pair these pieces together in all
replicas.  Thus, the inter-replica attraction can potentially force
the different replicas to ``lock'' together, i.e., to behave in an
correlated way. Indeed, the distribution of inter-replica correlations, 
usually measured
in terms of ``overlaps'', is a common device used to detect the
existence of a glass phase in disordered
systems~\cite{meza86}.

The full $n$-replica problem is difficult to solve analytically. We
will examine this problem in the regime of small $D$, aiming to
resolve the relevancy of disorder in a perturbative sense. Since the
lowest order term of the fully random problem in a perturbation
expansion in $D$ corresponds to the two-replica ($n=2$) problem, we
will focus on the latter in order to study the small-$D$ behavior of
the full problem. The solution of the two-replica problem will also
illustrate explicitly the type of interaction one is dealing with,
thereby providing some intuition needed to tackle the full problem.
It turns out that the two-replica problem can be solved exactly.
Here, we outline the saline features of the solution. Details of the
calculation and analysis are provided in the Appendices.  We will find
that the two-replica system has a phase transition between the molten
phase in which the two replicas are uncorrelated and a nontrivial phase
in which the two replicas are completely locked together in the
thermodynamic limit.  The transition occurs at a finite temperature
$T_{\mathrm{c}}(D)$ which approaches zero as $D\to0$. Thus, the effect of weak
disorder is {\em irrelevant} at finite temperatures.

\begin{figure}[htpb]
\begin{center}
\includegraphics[width=1.0\columnwidth]{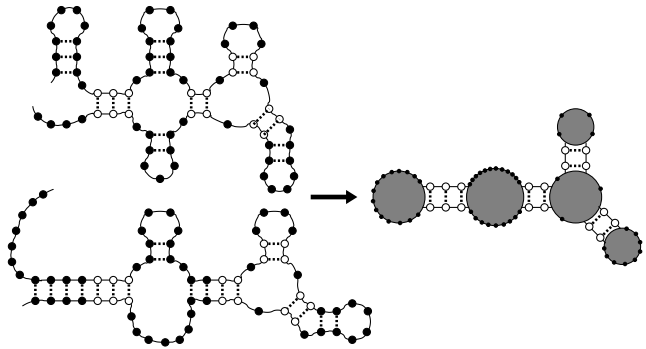}
\caption{Grouping of two RNA structures according to their common
bonds: Each pair of RNA secondary structures like the one on the
left hand side can be classified according to the bonds which are
common to both structures (open circles.) These common bonds form by
themselves an RNA secondary structure (right hand side.) Thus, the sum
over all pairs of secondary structures can be written as the sum over
all possible secondary structures of the common bonds. The weight of
each common bond structure is then given by the interaction energies
of common bonds and the summation over all possibilities of arranging
non-common bonds in the given common-bond structure. Since non-common
bonds have to be compatible with the common bond structure, the latter
sum factorizes into independent contributions of all the loops of the
common bond structure (grey circles.) Each such contribution solely
depends on the number of non-common-bond bases in each of these
loops.}
\label{fig_combonds}
\end{center}
\end{figure}

Let us denote the two-replica partition function $\overline{Z^2}$ 
for two strands each of length $N$ by
$\tworG(N+1;\widetilde{q})$, where we keep the dependence on $q$ implicit.
Then, 
\begin{equation}
\tworG(N+1;\widetilde{q}) = \sum_{S_1,S_2 \in \structures(N)} q^{|S_1|+|S_2|} 
\,\widetilde{q}^{|S_1\cap S_2|}.\label{eq_defG}
\end{equation}
The key observation which allows us to solve the two-replica problem
is that for each given pair of secondary structures, the bonds shared by
two replicas (hereafter referred to as ``common bonds'') 
form a valid secondary structure by themselves (see
Fig.~\ref{fig_combonds}.) Thus, we can rearrange the summation over
the pairs of secondary structures in the following way: 
We first sum over all possible
secondary structures of the common bonds.  For a {\em given
configuration} of the common bonds, we then sum over the remaining 
possibilities of intra-replica base pairings for each replica,
with the constraint that no new common bonds are created.

Note that the common bonds partition the diagram into a number of
``bubbles''\footnote{The two ends of the sequence must also belong 
to a bubble if they are not common bonds},  
shown as the shaded regions in Fig.~\ref{fig_combonds}.
Due to the exclusion of pseudo-knots from the valid secondary 
structures, only bases belonging to the same bubble can be paired 
with each other. Thus, the two-replica partition function can be written as
\begin{equation}\label{eq_gintermsofloops}
\tworG(N+1;\widetilde{q})=\sum_{S\in \structures(N)}(q^2\widetilde{q})^{|S|}
\hspace*{-3mm}\prod_{\mbox{bubble $i$ of $S$}}\hspace*{-3mm}
\tworQ_i(\ell_i+1),
\end{equation}
where the factor $q^2 \widetilde{q}$ is the weight of each common bond,
and 
$\tworQ_i(\ell_i+1)$ is the sum of all possible intra-replica pairings of
of the $i^{\rm th}$ bubble of $\ell_i$ bases in $S$, with the restriction that
there are no common bonds.

It should be clear that $\tworQ_i$ neither depends on the number of stems 
branching out from the bubble $i$ nor
the positions of these stems relative to the bases within the
bubble. It depends on $S$ only through the number
of bases $\ell_i$ in the bubble
and is given by a single function $\tworQ$ independent of $i$.
This function can be written explicitly as
\begin{equation}
\tworQ(\ell+1)\equiv \sum_{\begin{array}{c}S_1,S_2\in\structures(\ell)
\\S_1\cap S_2=\emptyset\end{array}}
\hspace{-5mm}q^{|S_1|+|S_2|}.
\label{eq_defQ}
\end{equation} 

With Eqs.~(\ref{eq_gintermsofloops}) and (\ref{eq_defQ}), the two-replica
problem is reduced to an effective {\em single} homogeneous RNA problem, 
with an effective
Boltzmann weight $q^2\widetilde{q}$ for each pairing, and an effective
weight $\tworQ$ for each single stranded loop.
As described in Appendix~\ref{app_intuitive}, this problem becomes formally
analogous to that of an RNA in the vicinity of the denaturation transition,
with $\tworQ$ being the weight of a single polymer loop fluctuating in 
$6$-dimensional embedding space.  
The competition between pairing energy and the bubble entropy leads to 
a {\em phase transition} for the two-replica problem, analogous to the
denaturation transition for a single RNA.

The details of this transition are given in Appendix~\ref{app_solution}, where
the partition function (\ref{eq_gintermsofloops}) is solved exactly. 
The exact solution exploits the relation
\begin{equation}
\tworQ(N) = \tworG(N;\widetilde{q}=0) \label{eq_GQ}
\end{equation}
which follows from the definitions (\ref{eq_defG}) and (\ref{eq_defQ}),
and turns Eq.~(\ref{eq_gintermsofloops}) into a recursive equation for
$\tworG$. The solution is of the form 
\begin{equation}
\tworG(N;\widetilde{q}) \sim N^{-\zexpo} \zeta^N(q,\widetilde{q})
\label{eq_fullG}
\end{equation}
for large $N$,
with two different forms for $\zexpo$ and $\zeta$ depending on whether
$\widetilde{q}$ is above or below the critical value 
\begin{equation}
\label{qc}
\widetilde{q}_c =
1+\frac{1}{q^2\sum_{N=1}^\infty N G^2(N)(1+2\sqrt{q})^{-2(N-1)}}.
\end{equation}
Here $G(N)$ is the molten phase partition function, whose
large $N$ asymptotics is given by
Eq.~(\ref{eq_moltenz}) and whose values for small $N$ can be calculated
explicitly from the recursion Eq.~(\ref{eq_g0recursion}). Thus,
the actual value of $\widetilde{q}_c$ can be found for any given $q$.

For $\widetilde{q} < \widetilde{q}_c$, we have $\zexpo=3$ and
\begin{equation}\label{eq_moltenzinmaintext}
\zeta = (1+2\sqrt{q})^2 + q^2 (\widetilde{q}-1) \gatsing(q),
\end{equation}
where
\begin{equation}\label{eq_defginmaintext}
\gatsing(q) = \sum_{N=1}^\infty G^2(N)(1+2\sqrt{q})^{-2N},
\end{equation}
according to Eqs.~(\ref{eq_gatsingdef}) and~(\ref{eq_moltenzresult}).
In this regime, the two-replica partition function $\tworG$ is
essentially a product of two single-replica partition functions $G$.
Compared to Eq.~(\ref{eq_moltenz}),
we can identify $\zexpo$ as $2\zexpo_0$, and $\zeta$ as a modified
version of $\singlezsing^2\equiv(1+2\sqrt{q})^2$. Since there is no
coupling of the two replicas beyond a trivial shift in the free energy
per length, $ f = -k_BT \ln\zeta$, we conclude that the disorder
coupling is irrelevant.  Hence the two-replica system is in the molten
phase in this regime.

For $\widetilde{q} > \widetilde{q}_c$, we have $\zexpo = 3/2$ and
$\zeta$ is given as the implicit solution of an equation involving
only single-replica partition functions as shown in
Eqs.~(\ref{eq_zkdef}) and~(\ref{eq_zetaresult}). Here, the partition
function of the two-replica system is found to have the {\em same}
form as that of the single-replica system in (\ref{eq_moltenz}). This
result implies that the two replicas are {\em locked} together via the
disorder coupling, and the molten phase is no longer applicable in
this regime.

Of course, as already explained above, only the weak-disorder limit
(i.e., $\beta^2D\ll 1$) 
of the two-replica problem is of relevance to the full random RNA problem.
In this limit,  $\widetilde{q} \approx 1 + \beta^2D$ while
$\widetilde{q}_c$ is found by evaluating Eq.~(\ref{qc}) with 
$q\approx e^{-\beta\avereps}$. It can be easily verified that
$\widetilde{q}_c > 1$ as long as $q$ is finite. Thus in the weak disorder
limit, we have $\widetilde{q} < \widetilde{q}_c$, indicating that
the molten phase is an appropriate description for the random
RNA. Unfortunately, the two-replica calculation cannot be used in
itself to deduce whether the molten phase description breaks down at
sufficiently strong disorder or low temperature. Based on this
analysis, we cannot conclude whether the type of phase transition
obtained for the two-replica problem is present in the full problem.

\section{Effect of Sequence Randomness: Low Temperature
Behavior}\label{sec_randomnesslowT}

Having established the validity of the molten phase description of
random RNA molecules at weak disorder or high temperatures, we now
turn our focus onto the low temperature regime. First, we will give
an analytical argument for the existence of a glass phase at
low temperatures. Then, we will present extensive numerical studies
confirming this result and characterizing this glass phase.

\subsection{Existence of a glass phase}\label{sec_proofglass}

We will start by showing that the molten phase cannot persist for all
temperatures down to $T=0^+$. To this end, we will {\em assume} that long
random RNA is in the molten phase for {\em all} temperatures, i.e., that
the partition function for any substrand of large length
$L\gg1$ is given by
\begin{equation}\label{eq_moltenzeff}
Z(L)=A(T)L^{-\frac32}\exp[-\beta f_0(T)L]
\end{equation}
with some effective temperature-dependent prefactor $A(T)$ and free
energy per length $f_0(T)$.  Then, we will
show that this assumption leads to a contradiction below some
temperature $T^*>0$. This contradiction implies that the molten phase
description breaks down at some finite $T_{\mathrm{g}}\ge T^*$. To be specific,
we will consider the sequence disorder model (\ref{eq_energychoice})
in this analysis.

The quantity we will focus on is again the free energy $\Delta
F(N)$ of the largest possible pinch. Under the assumption that
the random sequences are described by the molten phase, it is given by
\begin{equation}\label{eq_assumedmoltenpinch}
\overline{\Delta F}(N)=\frac32k_B T \ln N
\end{equation}
for large $N$ and all $T$ {\em independently} of the values of the effective
prefactor $A(T)$ and the free energy per length
$f_0(T)$ (see Eq.~(\ref{eq_moltenpinch}).)

\begin{figure}
\begin{center}
\includegraphics[angle=0,width=0.99\columnwidth]{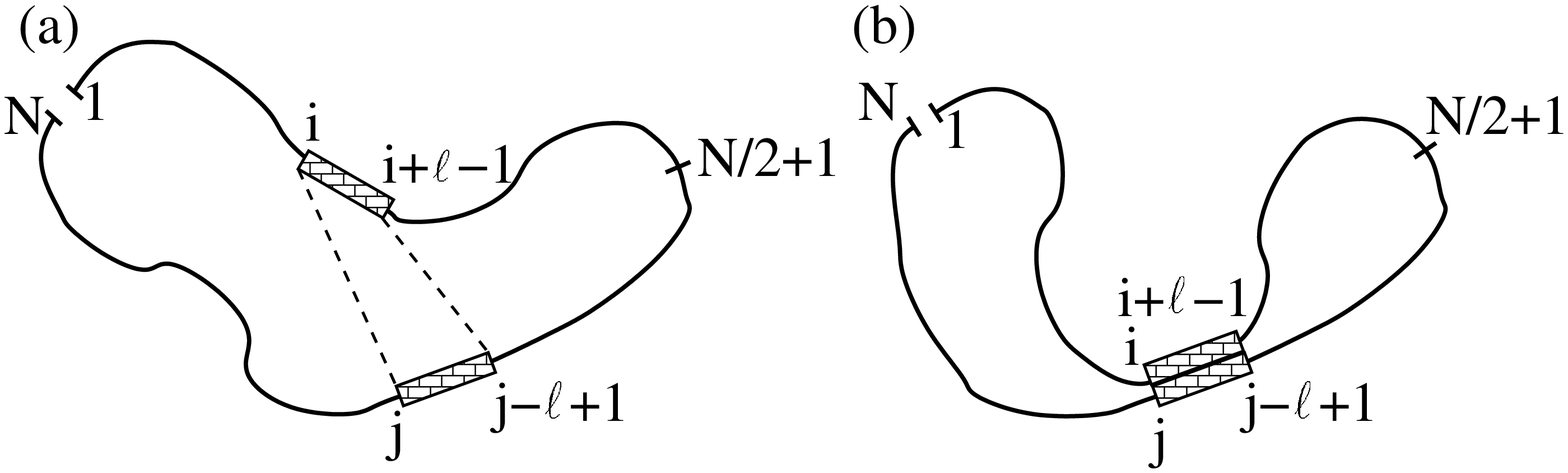}
\caption{Finding a good match in a RNA sequence: (a) shows the
position of two pieces with exactly complementary bases one of which
is between positions \protect$2$ and \protect$N/2$ and the other of
which is between positions \protect$N/2+2$ and \protect$N$. Such a
piece of length \protect$\ell\sim\ln N$ can be found for almost all
sequences.  (b) shows how restricting configurations to those in which
the good match forms Watson-Crick base pairs splits the molecule into
two loops which can still form base pairs within the loops
independently from each other.}
\label{fig_goodmatch}
\end{center}
\end{figure}

On the other hand, we can study this pinching free energy for each
given sequence of bases drawn from the ensemble of random sequences.
For each such sequence, we can look for a continuous segment of
$\ell\ll N$ Watson-Crick pairs
$b_i$--$b_j,b_{i+1}$--$b_{j-1},\ldots,b_{i+\ell-1}$--$b_{j-\ell+1}$
where the bases $b_i\ldots b_{i+\ell-1}$ are within the first half of the
molecule and the bases $b_{j-\ell+1}\ldots b_j$ are in the second half
(see Fig.~\ref{fig_goodmatch}(a).) For random sequences, the
probability of finding such exceptional segments decreases
exponentially with the length $\ell$, with the
largest $\ell$ in a sequence of length $N$ being typically 
\begin{equation}
\ell = \lambda^{-1} \ln N. \label{eq_lambda}
\end{equation}
For exact complementary matches, the proportionality constant
is known to be $\lambda = \ln 2$~\cite{arra90}.

Now, we calculate the pinching free energy
\begin{equation}\label{eq_deltafasdifference}
\Delta F(N)=F_{\mathrm{pinched}}-F_{\mathrm{unpinched}}
\end{equation}
by evaluating the two terms separately. The partition function
for the unpinched sequence contains {\em at least} all
the configurations in which the two complementary segments $b_i\ldots
b_{i+\ell-1}$ and $b_{j-\ell+1}\ldots b_j$ are completely paired (see
Fig.~\ref{fig_goodmatch}(b)). Thus,
\begin{equation}\label{eq_unpinchedandpaired}
F_{\mathrm{unpinched}}\le F_{\mathrm{paired}}
\end{equation}
where $F_{\mathrm{paired}}$ is the free energy of the ensemble of
structures in which the two complementary segments are paired. The
latter is the sum of the free energy of the paired segments and those
of the two remaining substrands $b_{i+\ell}\ldots b_{j-\ell}$ of
length $L_1\!=\!j\!-\!i\!-\!2\ell\!+\!1$ and $b_{j+1}\ldots b_{i-1}$
(wrapping around the end of the molecule) of length
$L_2\!=\!N\!+\!i\!-\!j\!-\!1$, i.e.,
\begin{equation}\label{eq_fpaired}
F_{\mathrm{paired}}\!\!=\!\!-\ell u_{\mathrm{m}} \!+ (N\!-\!2\ell) f_0
+{\textstyle\frac{3}{2}}k_B T\big[\!\ln(L_1)\!
+\!\ln(L_2)\big].
\end{equation}

The free energy $F_{\mathrm{pinched}}$ in the presence of the pinch is,
by the assumption of the molten phase, the interaction energy of the
pinched base pair $b_1$--$b_{N/2+1}$ plus the molten free energy of the
substrand $b_2\ldots b_{N/2}$ and the molten free energy of the
substrand $b_{N/2+2}\ldots b_N$, i.e., according to
Eq.~(\ref{eq_moltenzeff})
\begin{equation}\label{eq_fpinched}
F_{\mathrm{pinched}}=f_0(T)N+2 \times \frac{3}{2}k_B T\ln N
\end{equation}
up to terms independent of $N$. Combining this with
Eqs.~(\ref{eq_deltafasdifference}), (\ref{eq_unpinchedandpaired}),
and~(\ref{eq_fpaired}), we get
\begin{equation}
\Delta F(N)\!\ge\!
\frac{3}{2}k_B T[2\ln N\!-\!\ln L_1\!-\!\ln L_2]+
\ell [u_{\mathrm{m}}\!+\!2f_0(T)].
\end{equation}
Using the result (\ref{eq_lambda})  and that $L_1$
and $L_2$ are typically proportional to $N$, we finally obtain
\begin{equation}
\overline{\Delta F}(N)\ge [u_{\mathrm{m}}+2f_0(T)] \lambda^{-1} \ln N
\end{equation}
for large $N$. This is only consistent with
Eq.~(\ref{eq_assumedmoltenpinch}) if
\begin{equation}\label{eq_consistencycondition}
\frac{3}{2}k_B T\ge \lambda^{-1}[u_{\mathrm{m}}+2f_0(T)].
\end{equation}

Now, $f_0(T)$ is a free energy and is hence a monotonically decreasing
function of the temperature. Thus the validity of the inequality
(\ref{eq_assumedmoltenpinch}) depends on the behavior of its right
hand side at low temperatures. As $T\to 0$, the inequality can only
hold if $\sigma \equiv u_{\mathrm m}+2f_0(T=0) \le 0$.  Since the
average total energy at $T=0$ is $u_{\mathrm m}$ times the average
total number of matched pairs of a random sequence, then $2f_0(0)$ is
simply the fraction of matches and $\sigma$ is the fraction of bases
not matched (for $u_{\mathrm m} = 1$.) Clearly, $\sigma$ cannot be
negative, and the inequality (\ref{eq_assumedmoltenpinch}) must fail
at some finite temperature unless $\sigma=0$.

We can make a simple combinatorial argument to show 
that in most cases the fraction
$\sigma$ of unbound bases must strictly be  positive. To illustrate this,
let us generalize the ``alphabet size'' of the sequence disorder model of 
Sec.~\ref{sec_intenergies}  from $4$ to an arbitrary even integer $\alphsize \ge 2$. We will 
still adopt the energy rule (\ref{eq_energychoice}) where each
of the $\alphsize$  bases can form a ``Watson-Crick'' 
pair exclusively with one other base. Let us estimate the number 
of possible sequences for which the fraction of unmatched bases $\sigma$ 
is zero in  the limit of long sequence length $N$ at $T=0$. Since at $T=0$, 
only Watson-Crick (W-C) pairs can be formed, we only need to count the number 
of sequences for which the fraction of W-C paired bases is $1$. 
This means that except for a sub-extensive number
of bases, all have to be W-C paired to each other.
From the mountain picture (Fig.~\ref{fig_strreps}), it is
clear that the number of possible secondary structures for such sequences
must scale like $2^N$, since the fraction of horizontal steps is 
non-extensive so that at each step,
there are only the possibilities for the mountain to go up or down.
For each of the $N/2$ pairings in one of these $2^N$ structures, 
there are $\alphsize$ ways of choosing
the bases to satisfy the pairing. So for each structure, there are
$\alphsize^{N/2}$ ways of choosing the sequence that would guarantee
the structure. Since there are a total of $\alphsize^N$ sequences,
it is clear that the fraction of sequences with all (but a sub-extensive 
number of) W-C pairs becomes negligible if
\begin{equation}\label{eq_frustration}
(2\sqrt{\alphsize})^N<\alphsize^N.
\end{equation}
Thus, for $\alphsize\ge6$, we must have $\sigma>0$.

For $\alphsize=2$, the left hand side of
Eq.~(\ref{eq_frustration}) grows faster than its right hand side.
This reflects the absence of frustration in this simple two-letter  
model as already discussed at the end of
Sec.~\ref{sec_intenergies}. One way to retain frustration is to 
introduce additional constraints, e.g., the minimal hairpin length
used in Ref.~\cite{pagn00}. 
With this constraint, a
structure with a sub-extensive number of unmatched bases can only
contain a sub-extensive number of hairpins. In the mountain picture,
this means that except for a sub-extensive number of steps,  there is
only one choice to go up or down at every step. 
This changes Eq.~(\ref{eq_frustration}) to
$\alphsize^{N/2}<\alphsize^N$. It  ensures frustration since
$\sigma>0$ for {\em all} $\alphsize$. Since a minimal length of $3$
bases is necessary in the formation of a real hairpin, real RNA is certainly
frustrated by this argument.
The random sequence model which we study in
this paper is marginal since $\alphsize=4$ and there is no constraint on
the minimal hairpin length. In this case, all the prefactors on the two
sides of Eq.~(\ref{eq_frustration}) (e.g., the overcounting
of sequences that support more than one structure) must be taken into account.
We will not undertake this effort here, but will verify numerically in
Sec.~\ref{sec_estimatets} that  $\sigma>0$ also in this case.

In all cases with $\sigma>0$, it follows that there is
some unique temperature $T^*$ below which the consistency
condition~(\ref{eq_consistencycondition}) breaks down, implying the
inconsistency of the molten phase assumption in this regime.  From
this we conclude that there must be a phase transition away from the
molten phase at some critical temperature $T_{\mathrm{g}}\!\ge\!T^*\!>\!0$.
The precise value of the bound $T^*$ depends on $\lambda$ which in turn
depends on the stringency of the condition we impose on the rare matching 
segments. For instance, if we relax the condition of exact complementarity
between two segments to allow for matches within each segment, then 
the constant $\lambda$ will be reduced from $\ln 2$ and the value
of $T^*$ will increase. This will be discussed more in
Sec.~\ref{sec_estimatets}.

\subsection{Characterization of the glass phase}

The above argument does not provide any guidance on the properties of
the low temperature phase itself.
In order to characterize the statistics of
secondary structures formed at low temperatures,
we re-do the simulations reported in Sec.~\ref{sec_hightnumerics}
at $k_BT=0.025$ in energy units set by $u_{\mathrm{m}}=u_{\mathrm{mm}}=1$. 
At this temperature, an unbound base
pair is penalized with a factor $e^{40}$ relative to a Watson-Crick
base pair, and a non Watson-Crick base pair is penalized even
more. Thus, only the minimal energy structures contribute (for the
sequence lengths under consideration here), and we may regard this
effectively as at $T=0$. As in Sec.~\ref{sec_hightnumerics}, we
average over $10,000$ realizations of the disorder for
$N\in\{10,20,40,80,160,320\}$, over $2,000$ realizations for $N=640$ and over
$1,000$ realizations for $N\in\{1280,2560\}$.

\begin{figure}[ht]
\begin{center}
\includegraphics[angle=0,width=0.9\columnwidth]{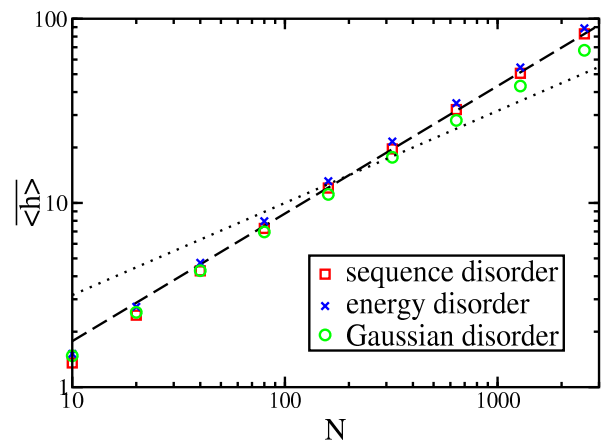}
\caption{Scaling of the average size \protect$\overline{\langle
h\rangle}$ of secondary structures in the low temperature phase with
the length \protect$N$ of the sequences: The plot shows data for three
different choices of disorder according to
Eqs.~(\protect\ref{eq_energychoice}),
(\protect\ref{eq_uncorrelateddisorder}),
and~(\protect\ref{eq_gaussiandisorder}) at \protect$T=0.025$. The
average system size follows a power law. However, the best fit of the
data for sequence disorder at \protect$N\ge160$ to a power law
indicated by the dashed line leads to an exponent of
\protect$\overline{\langle h\rangle}\sim N^{0.69}$. The corresponding
fits for energy and Gaussian disorder yield exponents of
\protect$\overline{\langle h\rangle}\sim N^{0.69}$ and
\protect$\overline{\langle h\rangle}\sim N^{0.65}$, respectively. This
is distinctively different from the square root behavior of the molten
phase indicated by the dotted line. The comparison of this plot with
its counterpart in Fig.~\protect\ref{fig_moltennumerics}(a) suggests
that the behavior of RNA secondary structures at low temperatures is
different from the molten phase.}
\label{fig_lowTnum}
\end{center}
\end{figure}

Fig.~\ref{fig_lowTnum} shows the results for
the ladder size $\langle h \rangle$ of the structures for the three
models of disorders.  The ladder size still scales algebraically
with the length of the sequences, with numerically determined
exponents ranging from $\overline{\langle h\rangle}\sim N^{0.65}$ to
$\overline{\langle h\rangle}\sim N^{0.69}$ for the different choices
of disorder. They are clearly different from the square root behavior
(dotted line)
expected of the molten phase. Thus this result reaffirms our expectation
 that the secondary structures of a random RNA sequence at zero temperature 
indeed belongs to a phase that
is different from the molten phase.

\subsubsection{A criterion for glassiness}

A key question in
characterizing the thermodynamic properties of disordered systems is
whether the zero-temperature behavior persists for a range of finite
temperatures.  If it does, then the system is said to have a
finite-temperature glass phase.  One way to address this question is
to study the overlap between different replicas of the RNA molecule as
mentioned earlier. If a non-trivial distribution of these overlaps
with significant weight on large overlaps persists into finite
temperatures, then the finite-temperature glass phase exists. This
approach was taken by previous numerical
studies~\cite{higg96,pagn00,hart01,pagn01}.  Unfortunately, the
results are inconclusive and even contradictory due to the 
weakness of the proposed phase transition --- only the fourth temperature
derivative of the free energy seems to show an appreciable
singularity.  Moreover, due to limitations in the sequence lengths probed,
it was difficult to get a good estimate of the asymptotic
behavior of the overlap distribution.

In our study, we adopt a different approach based on the droplet
theory of Fisher and Huse~\cite{huse91}. In this approach, one studies the
``large scale low energy excitations'' about the ground state.
This is usually accomplished by imposing a deformation over a length 
scale $\ell \gg 1$ and monitoring the minimal (free) energy cost
of the deformation. This cost is expected to scale as $\ell^\dropletexpo$
for large $\ell$.  A positive
exponent $\dropletexpo$ indicates that the deformation cost {\em
grows} with the size. If this is the case, the thermodynamics is dominated
by a few low (free) energy configurations in the thermodynamic limit,
  and the statistics of the zero-temperature
behavior persists into finite temperatures. 
On the other hand, if the exponent $\dropletexpo$ is negative, then there
are a large number of configurations which have low overlap with the ground
state but whose energies are similar to the ground state energy
in the thermodynamic limit. At any finite temperature $T$, a finite fraction
of these configurations (i.e., those within $O(k_BT)$ of the ground state
energy) will contribute to the thermodynamics of the system.
The zero temperature behavior is clearly not stable
to thermal fluctuations in this case, and no thermodynamic glass
phase can exist at any finite temperature. The analysis of the previous
section indicate the existence of a glass phase; thus we expect to find
the excitation energy to {\em increase} with the deformation size.

It should be noted that this criterion for
glassiness is purely thermodynamical in nature and does not make any
statement about {\em kinetics}. A system which is not glassy thermodynamically
can still exhibit very large {\em barriers} between the
many practically degenerate low energy configurations, leading to
a {\em kinetic glass}. A study of the kinetics of RNA, e.g., in terms
of barrier heights, is naturally dependent on the choice of allowed dynamical
pathways to transform one RNA secondary structure into another
one~\cite{morg96,morg98,flam00,isam00}. Since the latter is a highly
non-trivial problem, we will restrict ourselves to thermodynamics and
use the droplet picture explained above as our criterion for
the existence of a glass phase.

\subsubsection{Droplet excitations}

According to the criterion for glassiness just presented, our goal is
to determine the value of the exponent $\dropletexpo$ for random RNAs
numerically.  To this end, the choice of ``large scale low energy excitations''
needs some careful thoughts. As in every disordered system, there is a
very large number of structures which differ from the minimal energy
structure only by a few base pairs and which have an energy only
slightly higher than the minimum energy structure. These structures
are clearly  not of interest here. Instead
we need to find a controlled way of generating droplet excitations of various
sizes.

\begin{figure}[htbp]
\begin{center}
\includegraphics[angle=0,width=0.9\columnwidth]{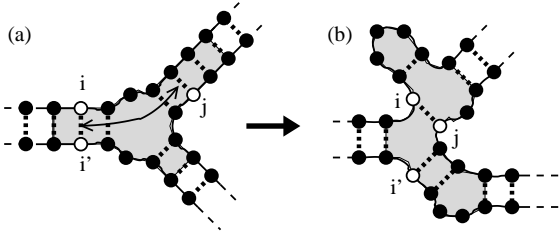}
\caption{Deformation of a minimal energy structure by
pinching: The two bases \protect$i$ and \protect$i'$ (open circles)
form a base pair in the indicated minimal energy structure (a).  Thus,
forcing these two bases to be bound by pinching does not affect the
structure at all.  Pinching of \protect$(i,j)$ on the other hand will
lead to a local rearrangement (shaded region)
of the structure as shown in (b). The
effect of such a pinch depends on the number  of
base pairs of the minimal energy structure the pinch is incompatible
with. As indicated by the arrow in (a), this number is given
by the ladder distance \protect$h_{i,j}$ 
between \protect$i$ and \protect$j$; in this example \protect$h_{i,j}=3$.}
\label{fig_pinchingstep}
\end{center}
\end{figure}

We propose to use the pinching method introduced in
Sec.~\ref{sec_observables} as a way to generate the deformation,
and regard the difference between the minimal energy pinched structure
and the ground state structure as the droplet excitation.
There are several desirable features 
about these pinch-induced deformations: First, it gives a convenient way 
of controlling the size of the deformation. If $(i,i')$ is a
base pair that is bound anyways in the ground state, pinching this
base pair does not have any effect and $\Delta F_{i,i'}=0$. If we
 pinch base $i$ with some other base $j\not= i'$, then we force 
at least a partial deformation of the ground state, for bases in the
vicinity of $i$, $i'$, and $j$. This is illustrated in
Fig.~\ref{fig_pinchingstep} with  the
deformed region indicated by the shade. As we move the pinch further away
from the ground state pairing, we systematically probe the effect of
larger and larger deformations (provided that a pinch only induces
{\em local} deformation as we will show). 
Second, the minimal energy or the free
energy of the secondary structures subject to the pinch constraint
is easily calculable numerically by the dynamic programming algorithm
as shown in Eqs.~(\ref{eq_returnprobability}) and~(\ref{eq_pinchfreeenergy}).
Third, the pinching of the bases in a sense mimics the actual dynamics
of the RNA molecule at low temperatures. In order for the molecule to
transform from one secondary structure to another at a temperature where
all matching bases should be paired, the bases have to make local 
rearrangements of the secondary structures much like the way depicted 
in Fig.~\ref{fig_pinchingstep}~\cite{flam00}. Thus, the pinching energy
provides the scale of variation in the local energy landscape for
such rearrangements\footnote{While local rearrangements will only proceed
by forming different Watson-Crick base pairs, we will in our
study determine the pinching free energies for {\em all} pinches
irrespective of the fact if they are a Watson-Crick base pair or not.
Since we take the ensemble average over many sequences this amounts
only to an irrelevant constant contribution
$\langle\epsilon_{ij}\rangle-u_{\mathrm{m}}=\avereps-u_{\mathrm{m}}$
to the pinch free energies.}.
Finally, ``pinching'' of a real RNA molecule can  
be realized in the pulling of a long molecule through 
a pore~\cite{gerl02}.

\begin{figure}[htbp]
\begin{center}
%\vspace*{7mm}
\includegraphics[angle=0,width=0.8\columnwidth]{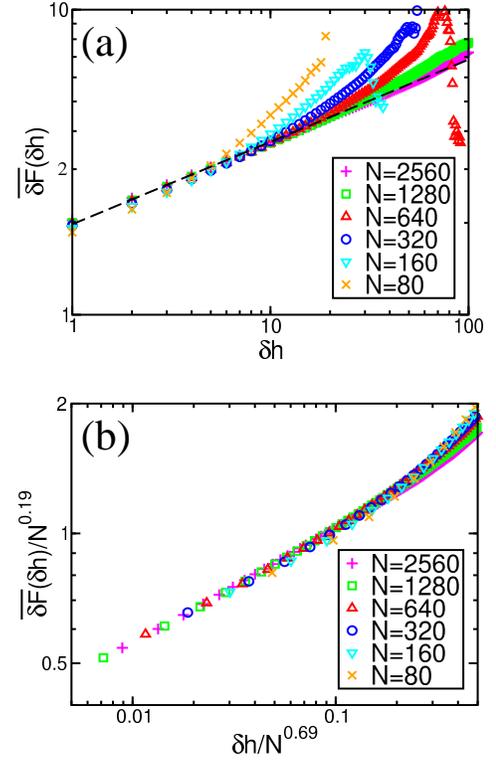}
\caption{Pinching free energy as a function of the number
\protect$\delta h$ of minimal energy structure base pairs that are
incompatible with the pinch for random sequence disorder at
\protect$T=0.025$: (a) shows the raw data. For small \protect$\delta
h$ the pinching free is independent of the length \protect$N$ of the
molecule and obeys a power law \protect$\overline{\delta F}(\delta
h)\sim(\delta h)^{0.27}$ (dashed line.) This is consistent with the
expectation that pinching at small \protect$\delta h$ leads to {\em
local} rearrangements of the secondary structure. The apparent
non-monotonic behavior at large \protect$\delta h$ is due to the small
number of sequences in which such a value of \protect$\delta h$ is
realized. (b) shows the same data, but the scaling of \protect$\delta
h$ with \protect$N$ is chosen in accordance with
Fig.~\ref{fig_lowTnum} while the scaling of the pinching free energy
is chosen in accordance with the power law dependence estimated
above.}
\label{fig_fofh}
\end{center}
\end{figure}

A key question to the utility of these pinch deformations is whether
the deformation is confined to the local region of the pinch as
depicted in Fig.~\ref{fig_pinchingstep} or whether it involves
a global rearrangement of the structure.  To test this aspect, we
numerically study the changes in pinch free energy as a function of
the ``size'' of a pinch. Here, the definition of the pinch size needs
some thought. Consider a specific sequence 
whose minimal energy structure is $S^*$.
If the binding partner of base $i$ is base $i'$ in the
minimal energy structure, a natural measure for the size of a
pinch $(i,j)\notin S^*$ with $i<j< i'$ would be the ladder distance $h_{i,j}$
between base $i$ and base $j$; see Fig.~\ref{fig_pinchingstep}. 
{From} the mountain representation (Fig.~\ref{fig_strreps}(b)), it
is easy to see that this is just the difference
 of the respective ladder distances of base $j$ and
base $i$ from base $1$ as defined in Eq.~(\ref{eq_defmountainheight}), i.e.,
$h_{i,j} = h_j(S^*)-h_i(S^*)$.
To find how the excitation energy depend on the pinch size, we just need
to follow how the pinching free energy $\Delta F_{i,j}$'s depend statistically 
on the size $h_{i,j}$'s. To do so, we choose a large number of random sequences,  
and determine the minimal energy structure $S^*$ for each of these sequences.  
Then, we compute the pinch free energies $\Delta F_{i,j}$ and the pinch size
$h_{i,j}$ for {\em all} possible pinches $(i,j)$ for each sequence. 
Afterwards, we average over all $\Delta F_{i,j}$'s with the same pinch
size $h_{i,j}$ over all of the generated random sequences
to obtain the function
\begin{equation}
\overline{{\delta F}}(\delta h)\equiv 
\sum_{i,j} \overline{\Delta F_{i,j} \, \delta_{\delta h,h_{i,j}}}
\Bigg/ \sum_{i,j} \overline{\delta_{\delta h,h_{i,j}}}.
\end{equation}
The results obtained at
$k_BT=0.025$ for a large range of sequence sizes from $N=80$ to $N=2,560$
are shown in Fig.~\ref{fig_fofh}(a). We see that the data for different
$N$'s fall on top of each other for small $\delta h$'s, with 
\begin{equation}
\overline{\delta F} \sim (\delta h)^{0.27}. \label{eq_Fh}
\end{equation}
This behavior explicitly shows that the pinch deformation is a {\em local}
deformation. In particular, we see that for small $\delta h$'s, 
the free energy cost is {\em independent} of the overall length $N$ of the
molecule. 

It is interesting to see at which $\delta h$ the entire sequence is involved.
One expects $\delta h_{\rm max} \sim \langle h \rangle \sim N^{0.69}$ 
since $\langle h \rangle$ gives the typical scale of the maximum ladder length.
To test this, we normalized $\delta h$ by $N^{0.69}$ and $\overline{\delta F}$
by $N^{0.19}$ (such that the relation (\ref{eq_Fh}) is preserved for small
$\delta h$'s). The result is shown in Fig.~\ref{fig_fofh}(b). We see that the
data is approximately collapsed onto a single curve, indicating that pinching
is indeed a good way of imposing a controlled deformation from the
ground state.

\subsubsection{A marginal glass phase}\label{sec_weakness}

The scaling plot of Fig.~\ref{fig_fofh}(b) indicates strongly that the energy
associated
with the pinch deformation {\em increases} with the size of the deformation, 
i.e., $\overline{\delta F} \sim (\delta h)^{0.27} \sim N^{0.19}$. However, 
the effective exponent involved is small, making the result very susceptible
to finite size effects. In order to decide on the glassiness of the system,
we want to focus on the energy scales associated with the largest pinch
deformations from the ground state. Assuming that there is only a single
energy scale associated with large pinches, we again study the 
free energy $\Delta F(N)$ of the largest  pinch as defined in
Eq.~(\ref{eq_largestpinchfreeenergy}) and average this over the ensemble
of sequences\footnote{In order to ensure that choosing the largest
pinch as a representative is justified, we studied in addition the
ensemble average of the {\em maximal} pinch free energy
$\Delta F_{\mathrm{max}}(N)\equiv\max_{1\le i<j\le N}\Delta F_{i,j}$.
This quantity yields  an
{\em upper bound} estimate of the energy associated with large scale
pinches for each sequence length $N$. We find 
$\overline{\Delta F_{\mathrm{max}}(N)}$ and $\overline{\Delta F}(N)$
to have the same scaling behavior, 
 and thus present only data for the latter.}.

\begin{figure}[htbp]
\begin{center}
\includegraphics[angle=0,width=0.95\columnwidth]{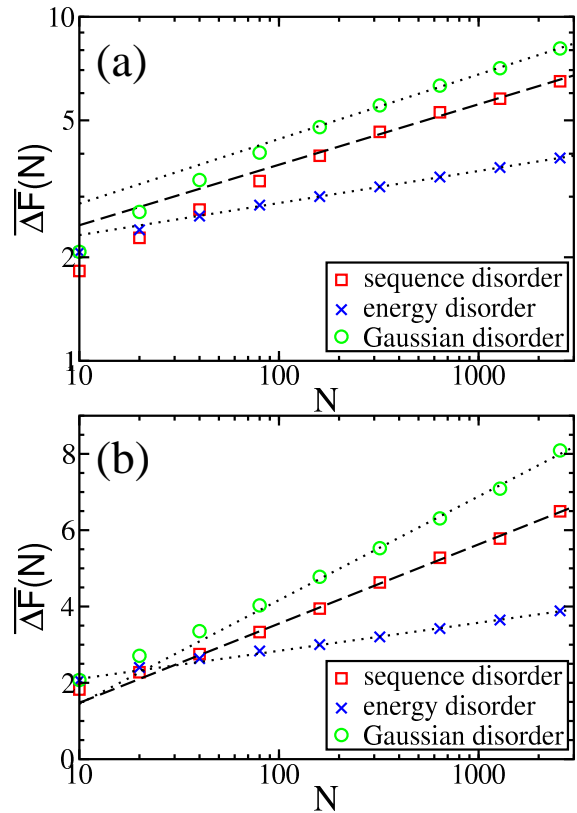}
\caption{Pinching free energies at low temperature: (a) shows a double
logarithmic plot with fits to power laws for the data with
\protect$N\ge160$. The exponents are \protect$N^{0.18}$,
\protect$N^{0.09}$, and $N^{0.19}$ for the sequence, energy, and
Gaussian disorder respectively. (a) shows the same data in a single
logarithmic plot together with the best fits to a logarithmic
dependence on \protect$N$. The statistical error of the data points is
about the size of the symbols for large \protect$N$ and smaller than
that for \protect$N\le640$. Due to the apparent systematic bending of
the data in the double logarithmic plot (b) we conclude that a
logarithmic dependence fits the data better although we cannot exclude
a power law behavior with a very small power.}
\label{fig_pinch}
\end{center}
\end{figure}

The results are shown in
Fig.~\ref{fig_pinch}(a) for the three models of disorders. 
Although a weak power law dependence of $\overline{\Delta F}(N)$ on $N$  
cannot be excluded, the fitted exponents obtained for the three models
are different from each other, ranging from $0.09$ to $0.19$.
This is a strong sign of concern, since the exponents are expected to
be independent of details of the models. In Fig.~\ref{fig_pinch}(b),
the same data
is plotted on a log-linear scale. The data fall reasonably on a straight
line for each of the models (especially for large $N$'s), 
suggesting  that the pinching free energy
may actually increase {\em logarithmically} with the sequence length,
similar to what is expected of the behavior in the  molten phase !
However in this case,  the prefactor of the logarithm depends on
the choice of the model and is much larger than the factor $\frac32 k_BT$
expected of the molten phase; see  Eq.~(\ref{eq_moltenpinch}).
For example, for the numerical data obtained at $k_BT=0.025$,
the prefactor  is approximately $0.9$ for the  sequence disorder model, 
while the expected slope for the molten phase is $0.0375$
at this temperature. Having different logarithmic prefactors for the 
different models is not a concern,
since a prefactor is a non universal quantity. Thus, our numerical results
favor a logarithmically increasing pinch energy, with a prefactor
much exceeding $k_BT$ at low temperature. 

What does this tell us about the possible glass phase of the random RNA? In
order to answer this question, we should remind ourselves that rather
special deformations are chosen in this study.  For our choice of
pinch deformations, we observe a logarithmic dependence of the gap
between the ground state energy and the energy of the excited
configurations on the length of the sequence or deformation. This
corresponds to the marginal case of the droplet theory where the
exponent $\dropletexpo$ vanishes. Since the pinching free energies are
increasing with length, we cannot exclude a glass phase in the case
$\dropletexpo=0$. We can say, though, that the increase of the
excitation energy with length is at most a power law with a very small
exponent and most probably even less than any power law. Therefore, a
possible glass phase of RNA has to be {\em very weak}. If it turns out
that the excitation energy is indeed a logarithmic function of length,
with a non-vanishing prefactor as $T\to 0$ as our numerics suggest,
then the low-temperature phase would be categorized formally as a {\em
marginal} glass phase, analogous to behaviors found in some
well-studied model of statistical mechanics~\cite{card82,kors93,hwa94}.
In any case, we should note that the actual
difference in the excitation energy is only a factor of $4$
across two-and-a-half decades in length. Thus the glassy effect will be
weak for practical purposes.  On the other hand, the weak dependence 
of the excitation energy on length may be the underlying
cause of discrepancies in the literature~\cite{pagn00,hart01,pagn01}
regarding the existence of the glass phase for the random RNA.

\subsection{Estimation of the phase transition
temperature}\label{sec_estimatets}

Now that we have studied in great detail the behavior of random RNA in
the low and the high temperature phase, we describe its behavior
at intermediate temperatures. To this end, we again study the
pinch free energies $\overline{\Delta F}(N)$ 
defined in Eq.~(\ref{eq_largestpinchfreeenergy}),
but this time over a large range of
temperatures. We concentrate on the sequence disorder model
Eq.~(\ref{eq_energychoice}) with $u_{\mathrm{m}}=u_{\mathrm{mm}}=1$, and 
study sequences of lengths up to $N=1,280$.

\begin{figure}[htbp]
\begin{center}
\includegraphics[width=1.0\columnwidth]{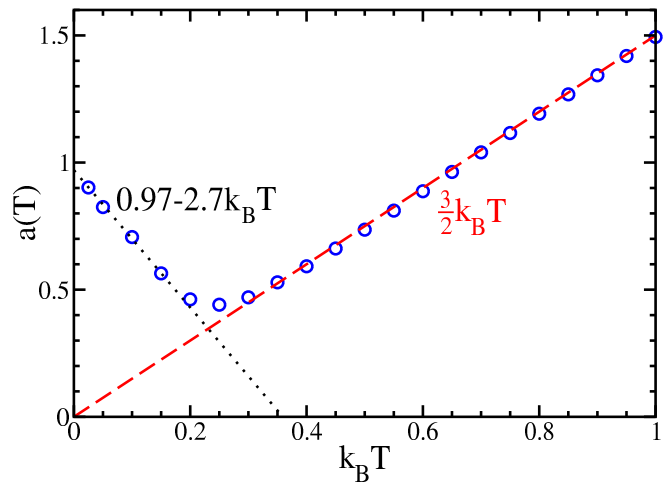}
\caption{Prefactor \protect$a(T)$ of the logarithmic dependence of 
$\overline{\Delta F}$ on $N$ for random RNA sequences generated by
the sequence disorder model: At
high temperatures, the prefactor indicated by the circle is well-described
by the dashed line \protect$\frac{3}{2}k_BT$ expected of the molten
phase. At low temperatures, it again has a linear temperature dependence
and is empirically fitted by the dotted line, $0.97 - 2.7k_BT$.
The numerical uncertainty in $a(T)$ is of the
order of or smaller than the size of the symbols.}
\label{fig_logfactors}
\end{center}
\end{figure}

{From} Secs.~\ref{sec_scaling},
\ref{sec_hightnumerics}, and~\ref{sec_weakness}, we know that the
pinch free energy $\overline{\Delta F}(N)$
depends logarithmically on the sequence length $N$ at both low
and high temperatures. Indeed, this logarithmic behavior seems
to hold for {\em all} temperatures studied.  The data for each temperature
can easily be fitted to the form
\begin{equation}\label{eq_logform}
\overline{\Delta F}(N)=a(T)\ln N + c(T)
\end{equation}
The prefactor $a(T)$ is found to depend on temperature in a
non-monotonic way as shown in Fig.~\ref{fig_logfactors}.
The figure contains values of $a(T)$ extracted by fitting the data
for $N\ge160$ to the form Eq.~(\ref{eq_logform}). The uncertainty of
this fit is on the order of the size of the symbols or smaller. 
For high temperatures,
we find $a(T) \approx  \frac{3}{2}k_BT$ (dashed line in
Fig.~\ref{fig_logfactors}) as expected for the molten phase. At
low temperatures, it starts from a finite value of the order $1$
and decreases linearly with temperature, as 
$a(T) \approx 0.97 - 2.7k_BT$ (dotted line in Fig.~\ref{fig_logfactors}).
If we identify the glass
transition temperature $T_g$ as the intersection of the dashed and the
dotted lines, we get
\begin{equation}\label{eq_tgestimate}
k_BT_{\mathrm{g}}\approx 0.25.
\end{equation}

It is interesting to compare this estimate with the lower bound $T^*$ for
the glass transition temperature given in Sec.~\ref{sec_proofglass}.
According to the consistency condition~(\ref{eq_consistencycondition}),
this lower bound is defined by
\begin{equation}\label{eq_defoftstar}
\lambda^{-1}[u_{\mathrm{m}}+2f_0(T^*)]=\frac{3}{2}k_B T^*
\end{equation}
with $\lambda=\ln2$. It is necessary to determine the
temperature dependence of the quantity $u_{\mathrm{m}}+2f_0(T)$ on the
left hand side of this equation numerically.  To do this, we measure the total
free energy of each sequence generated.
Averaging these free energies over all the sequences of a
given length $N$ and temperature $T$,  and
dividing the results by the respective lengths $N$, we obtain an estimate
$f_0(T;N)$ of the free energy per length which approaches the desired $f_0(T)$
for large $N$.

\begin{figure}[htbp]
\begin{center}
\includegraphics[angle=0,width=0.99\columnwidth]{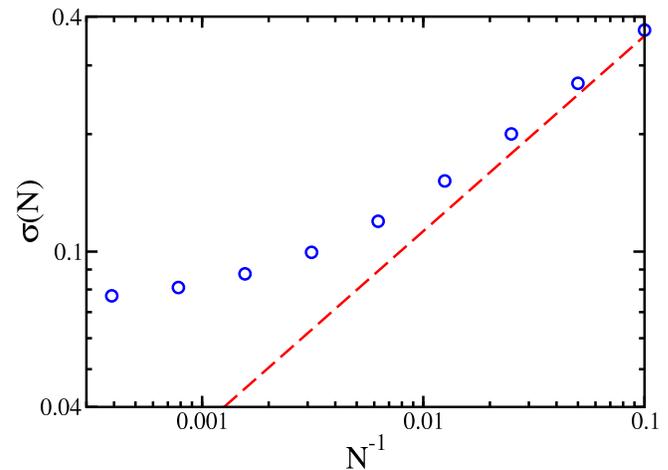}
\caption{Sequence length dependence of the fraction of unbound bases 
\protect$\sigma(N) =  u_{\mathrm{m}}+2f_0(T\approx0;N)$:
The data shown is taken at \protect$k_BT=0.025$. For small \protect$N$
it is dominated by the statistical fluctuations in the number of bases
according to Eq.~(\protect\ref{eq_smallnf0}) (the dashed line). At large
\protect$N$, it saturates to a positive constant.}
\label{fig_ump2f0ofN}
\end{center}
\end{figure}

Fig.~\ref{fig_ump2f0ofN} shows how
these estimates depend on the sequence length $N$ for the lowest
temperature $k_BT=0.025$ studied. Instead of the free energy per
length itself, the figure shows the fraction of unbound bases
$\sigma(N) =  u_{\mathrm{m}}+2f_0(T\approx0;N)$.
For short sequences these estimates show a clear dependence on the
sequence length $N$. This can be understood in terms of sequence-to-sequence
 fluctuations in the maximum number of possible pairings, due to fluctuations 
in the actual number of each type of bases present in a given sequence, even
if all four bases are drawn with equal probability. 
This effect can be quantified by assuming that there is
no frustration for small $N$, i.e., for any given sequence of the
four bases $A$, $C$, $G$, and $U$, a secondary structure with 
the maximal number of Watson-Crick base pairs can be formed. If
we denote by $n_X$ the number of times that the base $X$ appears in
the sequence, the maximal number $M$ of pairings is given by 
$M=\min\{n_A,n_U\}+\min\{n_G,n_C\}$. The fraction of unbound bases
$1-2M/N$
due to this effect can be computed straightforwardly approximating the
multinomial distribution of $n_A-n_U$ by an appropriate Gaussian distribution,
with the result
\begin{equation}\label{eq_smallnf0}
1-\frac{2M}{N}\approx2/\sqrt{\pi N}.
\end{equation}
We expect this effect to be responsible for the increase in $\sigma(N)$
found in Fig.~\ref{fig_ump2f0ofN}. Indeed this effect, as indicated by the
dashed line in the figure, explains the $N$ dependence of $\sigma(N)$ well for 
$N < 100$. However, we also see from the figure a clear saturation effect
at large $N$. This saturation reflects the finite
fraction of unbound bases which is a frustration effect
forced upon the system through the restriction on the type of allowed 
pairings in a allowed secondary structure. The unbound fraction 
$\sigma \approx 0.08$ is {\em finite} asymptotically as expected in
Sec.~\ref{sec_proofglass}. Note that this value is {\em
remarkably small}, as it implies that in the ground state structure of
our toy random sequence, more than $90\%$ of the maximally
possible base pairs are formed. But this is an artifact of the very
simple energy rule used in our toy model.  This fraction certainly will
become smaller if the realistic energy rules are used, making the 
system more frustrated hence more glassy.

\begin{figure}
\begin{center}
\includegraphics[angle=0,width=0.99\columnwidth]{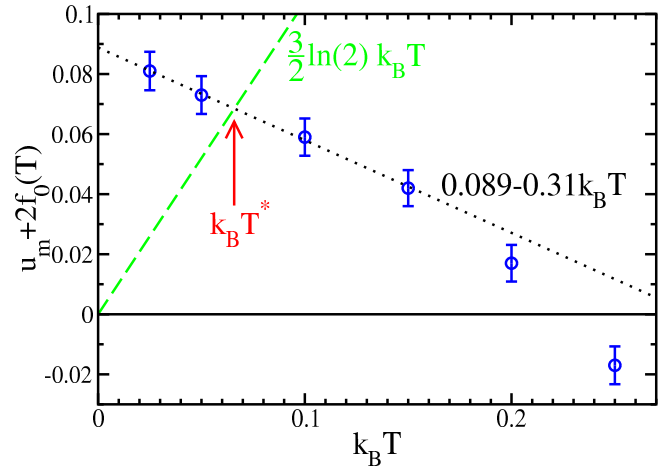}
\caption{Estimation of \protect$T^*$: The symbols show numerical
estimates of the quantity \protect$u_{\mathrm{m}}+2f_0(T)$ for
different temperatures. The estimates are obtained by averaging the
numerically determined free energy of \protect$1,000$ random sequences of
length \protect$N=1,280$ generated by the sequence disorder model
Eq.~(\protect\ref{eq_energychoice}) with
\protect$u_{\mathrm{m}}=u_{\mathrm{mm}}=1$. The low temperature behavior
can be described reasonably well by the expression 
\protect$0.089 - 0.31k_BT$ (dotted line).
The consistency
condition~(\protect\ref{eq_consistencycondition}) for the molten phase
breaks down when this line intersects
\protect$\frac{3}{2}\ln(2)k_BT$ (the dashed line).
This yields \protect$k_BT^*\approx0.066$ as a lower bound
for the glass transition temperature of this system.}
\label{fig_ump2f0}
\end{center}
\end{figure}

In order to obtain the temperature dependence of the quantity 
$u_{\mathrm{m}}+2f_0(T)$ on the  left hand side
of Eq.~(\ref{eq_defoftstar}), we will use its value at
$N=1,280$ as an estimate of its asymptotic value.
The results are shown in Fig.~\ref{fig_ump2f0}.
The behavior at low temperatures can be described by a linearly
decreasing function, shown as the dotted line in Fig.~\ref{fig_ump2f0}.
According to Eq.~(\ref{eq_defoftstar}), the temperature $T^*$
is obtained as the intersection of this curve and $\frac32\lambda k_BT$,
shown as the dashed line in Fig.~\ref{fig_ump2f0} for $\lambda = \ln 2$.
We find
\begin{equation}\label{eq_valueoftstar}
k_BT^*\approx0.066
\end{equation}
which is consistent with the estimate (\ref{eq_tgestimate}), but is
a rather weak bound.
Improved bounds on $T_{\mathrm{g}}$ can be made by relaxing the
condition of perfect complementarity of the two segments imposed in
Sec.~\ref{sec_proofglass}. This leads to larger values of the prefactor
$\lambda^{-1}$ in Eq.~(\ref{eq_defoftstar}), hence a smaller slope
for the dashed line in Fig.~\ref{fig_ump2f0}, and a larger value
of $T^*$. While the details of improved bounds will be
discussed elsewhere, let us remark here that from Fig.~\ref{fig_ump2f0}
it is clear  that no matter what the slope of the
dashed line becomes, we will never have $T^*$ larger than the temperature
of $k_B T \approx 0.22$ where the quantity $u_{\mathrm{m}}+2f_0(T)$ goes below zero.
Thus, these estimates will always be
consistent with the observed glass transition temperature of
$k_BT_g\approx0.25$. 

Moreover, we note that the low temperature behavior 
$u_{\mathrm{m}}+2f_0(T) \approx 0.089 - 0.31k_BT$ as indicated by the dotted
line in Fig.~\ref{fig_ump2f0}, appears to be roughly related to the behavior of
$a(T)$ (dotted line in Fig.~\ref{fig_logfactors}) in the same temperature
range, by a single scaling factor of approximately $0.1$. Thus, it is
possible that 
\begin{equation}
a(T) \approx \lambda^{-1} [u_{\mathrm{m}} + 2 f_0(T)]
\end{equation}
if it turns out that $\lambda^{-1} \approx 0.1$ for $T<T_g$.
If this is the case, then it means the procedure we used to estimate
the pinch energy in Sec.~\ref{sec_proofglass} is quantitatively correct,
implying that the ground state of a random RNA sequence indeed consists of the
matching of rare segments independently at each length scale.
It will be useful to pursue this analysis further using a renormalization
group approach similar to what was developed for the denaturation 
of heterogeneous DNAs by Tang and Chat\'e~\cite{tang00}.

\section{Summary and Outlook}

In this manuscript, we studied the statistical properties of random RNA sequences
far below the denaturation transition so that bases predominantly form 
base pairs. We introduced several toy energy models which allowed us
to perform detailed analytical and numerical studies. Through a
two-replica calculation, we show that sequence disorder is
perturbatively irrelevant, i.e., an RNA molecule with weak sequence
disorder is in a molten phase where many secondary structures with
comparable total energy coexist.  A numerical study of the model at
high temperature recovers scaling behaviors characteristic of the molten
phase. At very low temperatures, a scaling argument based on the extremal
statistics of rare matches suggest the existence of a different phase.
This is supported by extensive numerical results: Forced
deformations are introduced by pinching distant monomers along the backbone; 
the resulting excitation energies are found to grow very slowly (i.e.,
logarithmically) with the deformation size. It is likely that the
low temperature phase is a marginal glass phase. The intermediate 
temperature range is also studied numerically. The transition between
the low temperature glass phase and the high temperature molten phase
is revealed by a change in the coefficient of the logarithmic excitation 
energy, from being disorder dominated to entropy dominated.

{From} a theoretical perspective, it would be desirable to find an
analytical characterization of the low temperature phase. If the
excitation energy indeed diverges only logarithmically, one has the
hope that this may be possible, e.g., via the replica theory, as
was done for another well known model of statistical physics~\cite{kors93}. It
should also be interesting to include the spatial degrees of freedom
of the polymer backbone (via the logarithmic loop energy), to see how
sequence disorder affects the denaturation transition. Another
direction is to include sequence {\em design} which biases a specific
secondary structure, e.g., a stem-loop~\cite{bund99}.  From a
numerical point of view, it is necessary to perform simulations with
realistic energy parameters to assess the relevant temperature regimes
and length scales where the glassy effect takes hold. To make potential
contact with biology, one needs to find out 
whether a molten phase indeed exists between the high temperature
denatured phase and the low temperature glass phase for a real random
RNA molecule, and which phase the molecule is in under normal physiological
condition.  Finally, it will
be very important to perform {\em kinetic} studies to explore the
dynamical aspects of the glass phase.  Despite the apparent weakness
of the thermodynamic glassiness, the kinetics at biologically relevant
temperatures is expected to be very slow for random sequences~\cite{isam01}.

\begin{acknowledgments}
The authors benefitted from helpful discussions with U.~Gerland and
D.~Moroz. T.H.\ acknowledges an earlier collaboration with D.~Cule
which initiated this study, and is indebted to L.-H. Tang for a
stimulating discussion during which the simple picture of
Sec.~\ref{sec_proofglass} emerged.  This work is supported by the
National Science Foundation through grants no.\ DMR-9971456 and
DBI-9970199. The authors are grateful to the hospitality of the
Institute for Theoretical Physics at UC Santa Barbara where this work
was completed.

\end{acknowledgments}

\appendix

\section{Heuristic derivation of the two-replica phase
transition}\label{app_intuitive}

Before we describe the exact solution for the two-replica problem, as
defined by the partition function $\tworG$ in Eq.~(\ref{eq_gintermsofloops})
and the bubble weight $\tworQ$ in Eq.~(\ref{eq_defQ}),
we first provide here a heuristic derivation of the qualitative results.
This mainly serves to give a flavor of the two-replica problem
in the language of theoretical physics.

To this end, we define the quantity $\Pi(N)$ to be the partition
function over all two replica configurations of a sequence of length
$N-1$ under the constraint that base $1$ and base $N-1$ form a common
bond. It is easy to see that
\begin{equation}\label{eq_gandpi}
\Pi(N)=\qhat\,\tworG(N-2;\widetilde{q})
\end{equation}
where we set
\begin{equation}
\qhat\equiv q^2\widetilde{q}.
\end{equation}
Thus, the critical behavior of $\tworG(N;\widetilde{q})$ is identical
to the critical behavior of $\Pi(N)$ which we will study
in the following.

\begin{figure}[htbp]
\begin{center}
\includegraphics[angle=0,width=1.0\columnwidth]{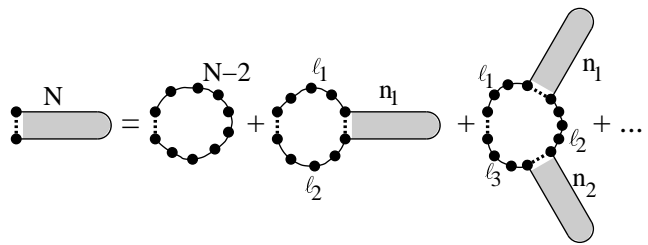}
\caption{Recursion equation for the restricted partition function
\protect$\Pi(N+1)$: While the first and last of the \protect$N$ bases
described by \protect$\Pi(N+1)$ always form a base pair this
outermost base pair can be followed by a loop with \protect$0,1,2,\ldots$
outgoing stems. Each of the stems is described by \protect$\Pi$ itself
while the loop is characterized by its total length that can be split
into the different pieces \protect$\ell_i$ in different ways.}
\label{fig_pirecurs}
\end{center}
\end{figure}

Due to the no pseudo knot constraint of the secondary structures, 
$\Pi(N)$ has a very simple structure, 
\begin{eqnarray}\label{eq_Pi.1}
\lefteqn{\Pi(N\!+\!1)=\qhat \tworQ(N\!-\!1)}
\hspace*{12mm}&&\\\nonumber
&&+\!\!\!\!\sum_{\ell_1,\ell_2,n_1}\!\!\!\qhat
\tworQ(\ell_1\!+\!\ell_2\!+\!1) \Pi(n_1\!+\!1) 
\delta_{\ell_1+\ell_2+n_1,N-2}\\\nonumber
&&+\!\!\!\!\sum_{\ell_1,\ell_2,\ell_3,n_1,n_2}\!\!\!\!\!\!\!\qhat
\tworQ(\ell_1\!+\!\ell_2\!+\!\ell_3\!+\!1)\Pi(n_1\!+\!1)\times\\\nonumber
&&\qquad\qquad\times
\Pi(n_2\!+\!1)\delta_{\ell_1+\ell_2+\ell_3+n_1+n_2,N-2}\!+
\ldots 
\end{eqnarray}
as illustrated in Fig.~\ref{fig_pirecurs}.
To simplify the above equation, it is useful to introduce the $z$-transforms
\begin{eqnarray*}
\Pihat[\mu]&=&\sum_{N=1}^\infty \Pi(N) e^{-\mu N}\\
\Qhat[y]&=&\sum_{N=1}^\infty \tworQ(N) e^{-yN}
\end{eqnarray*} 
of $\Pi$ and $\tworQ$.
Now applying the $z$-transform to both sides of Eq.~(\ref{eq_Pi.1}), we 
obtain
\begin{eqnarray}
\Pihat[\mu]e^{2\mu}\!\!\!&=&\!\!\!
\qhat\!\! \int\!\! \frac{\mathrm{d}y}{2\pi i} \Qhat[y] \Khat[y,\mu]
\Big[ 1\! +\! \Khat[y,\mu]\Pihat[\mu]e^{2\mu}e^{-y} \label{eq_Pi.2}\\\nonumber
&&\hspace*{25mm} +\! \left(\Khat[y,\mu]\Pihat[\mu]e^{2\mu}e^{-y}\right)^2
\!\!\!\!+\ldots \Big]
\end{eqnarray}
where 
\begin{equation}
\Khat[y,\mu] = \frac{1}{e^{-y+\mu}-1},
\end{equation}
and the inverse transform
$\tworQ(\ell) = \int \frac{\mathrm{d}y}{2\pi i} \Qhat[y] e^{y\ell}$ was used.

Eq.~(\ref{eq_Pi.2}) can be simplified greatly to the following form, 
\begin{equation}
\Pihat[\mu]e^{2\mu} = \qhat\cdot
\int\frac{\mathrm{d}y}{2\pi i}
\frac{\Qhat[y]}{\Khat^{-1}[y,\mu]-\Pihat[\mu]e^{2\mu}e^{-y}}.
\label{eq_Pi}
\end{equation}
This is reminiscent of the well-known Hartree solution to
the $\phi^4$-theory, or equivalently the self-consistent treatment of the
self-interacting polymer
problem~\cite{zinn89}, if we identify $\qhat$
as the interaction parameter, $\Khat[y,\mu]$ as the ``propagator".
The usual form of the Hartree equation
\begin{equation}
\Pihat[\mu] = \qhat\cdot
\int \frac{\mathrm{d}^dk}{(2\pi)^d} \, \frac{1}{k^2 + \mu -\Pihat[\mu]}
\label{eq_Pi1}
\end{equation}
corresponds to the small-$y$,
small-$\mu$ limit of Eq.~(\ref{eq_Pi}), 
with $-y$ playing the role of the square of the ``wave
number" $k$. Note that  $\Qhat[y]$ plays the role of 
the density of (spatial) states,
i.e. $dy \Qhat[y] = \frac{\mathrm{d}^dk}{(2\pi)^d}$, where $d$
denotes the {\em dimensionality} of the ``embedding space".

In the context of RNA, de Gennes used this approach to describe the
denaturation of uniformly attracting RNA more than 30 years ago~\cite{dege68}.
Recently, this approach has been extended by Moroz and Hwa to study
the phase diagram of RNA structure formation~\cite{moro01}. The analysis of a
self-consistent equation of the type (\ref{eq_Pi}) is well known~\cite{moro01,zinn89}.
The analytical properties of $\Pihat[\mu]$ depend crucially on the 
form of $\Qhat[y]$. Let the singular part of $\Qhat$ be 
\begin{equation}
\Qhat_{\mathrm{sing}} \propto (y-y_{\mathrm{c}})^{\alpha-1}\label{eq_Qsing}
\end{equation}
where $y_{\mathrm{c}}$ is the position of the singularity of
$\Qhat[y]$. [Note that $\alpha= d/2$ by comparing the forms of the Hartree
equations (\ref{eq_Pi}) and (\ref{eq_Pi1}).]  For $1<\alpha < 2$, there is only one solution for all
$\qhat>0$, with a square root singularity in $\Pihat[\mu]$ at some
finite value of $\mu$.  For $\alpha > 2$, there are two possible
solutions depending on the value of $\qhat$. The square-root
singularity exists for $\qhat$ exceeding some critical
value\footnote{Note that the critical value
\protect$\qhat_{\mathrm{c}}(q)$ depends through \protect$\Qhat$ on
\protect$q$ but not on \protect$\widetilde{q}$.}
$\qhat_{\mathrm{c}}(q)>0$, while for $\qhat<\qhat_{\mathrm{c}}(q)$,
the square-root singularity disappears and $\Pihat[\mu]$ is governed
by the singularity of $\Qhat$ given in Eq.~(\ref{eq_Qsing}).
Performing the inverse transform and using Eq.~(\ref{eq_gandpi}), we
get $\tworG(N;\widetilde{q}) \sim N^{-\thetahat} \zeta^N$ where
$\zeta$ is a non-universal parameter given by the location of the
singularity, while the exponent $\thetahat$ characterizes the phase of
the system and is given by the singularity of $\Pihat[\mu]$: We have
$\thetahat = 3/2$ if $\Pihat[\mu]$ is dominated by the square-root
singularity and $\thetahat = \alpha$ if $\Pihat[\mu]$ is dominated by
$\Qhat$.

The interpretation of the two phases with $\thetahat = 3/2$
and $\thetahat = \alpha$ are straightforward: The phase with $\thetahat=3/2$
describes the usual RNA secondary structure (see Eq.~(\ref{eq_moltenz}));
here the 
bubbles described by $\tworQ$ are irrelevant. In the other phase, 
the result that $\thetahat = \alpha$ indicates that base pairing is
not relevant and the system behaves as a single bubble.
In the context of the original two-replica problem, the irrelevancy
of the bubbles in the $\thetahat=3/2$ phase indicates that the two
replicas are locked together, behaving as a single replica in this phase.
In the other phase, the attraction of the common bonds is irrelevant,
and the two replicas become independent of each other.

As explained in Sec.~\ref{sec_replica}, the purpose of the two-replica
calculation is to determine whether the inter-replica attraction,
characterized by $\qhat$ here, is irrelevant, i.e., whether the system
will not yet be in the $\thetahat=3/2$ phase for a value of
$\qhat\gtrsim q^2\!\cdot\!1$. This is only possible if
$\qhat_{\mathrm{c}}(q) > 0$. From the solution of the problem
described above, this depends crucially on the singularity of $\Qhat$,
specifically, on whether $\alpha>2$.  The difficulty in ascertaining
the form of $\Qhat$ lies in the no-common-bond constraint (i.e., $S_1
\cap S_2 = \emptyset$) in the definition of $\tworQ$ (\ref{eq_defQ}). However, we note
that for $\widetilde{q}=1$, the two replica partition function
$\tworG(N;\widetilde{q}=1)$ is simply the square of the single replica
partition function $G(N)$.  Thus, $\tworG(N;\widetilde{q}=1) = G^2(N)
\sim N^{-2\zexpo_0} \singlezsing^{2N}(q)$ according to
Eq.~(\ref{eq_moltenz}). Since we just convinced ourselves that
$\thetahat$ can take on only two possible values, namely $\alpha$ and
$3/2$ and since $2\zexpo_0=3\not=3/2$ we conclude
$\alpha=2\zexpo_0=3>2$ and moreover
$q^2\le\qhat_{\mathrm{c}}(q)$. Thus, we do expect the phase transition
to occur at $\qhat_{\mathrm{c}}(q) >0$.  However, it is not clear from
this calculation if the system at $\widetilde{q}=1$ (or $\qhat=q^2$) is
exactly at or strictly below the phase transition point. We leave it
to the exact solution of the two replica problem presented in the next
two appendices to establish that $\qhat=q^2$ is indeed strictly
below the phase transition point and that therefore disorder is
perturbatively irrelevant.

We note that in the context of the $\phi^4$-theory or the self-consistent
treatment of the self-interacting polymer, the result $\alpha=3$ implies
that the embedding spatial dimension is $d=6$. Thus, the two-replica problem
corresponds to the denaturation of a single RNA in $6$ spatial dimensions.
The bubbles $\tworQ$'s of Fig.~(\ref{fig_combonds}) which originate from 
the branching entropy of the individual RNAs play the role of spatial 
configurational entropy of the single-stranded RNA in the denaturation problem.

\section{Solution of the two-replica problem}\label{app_solution}

In this appendix, we present the exact solution of the two replica
problem. While most of the details are given here, the most laborious
part is further relegated to App.~\ref{app_derivez1}.

\subsection{An implicit equation for the two-replica
problem}\label{app_tworeplicarecursion}

We start by introducing an auxiliary quantity
$\tworW(N,n;\widetilde{q})$. This is a restricted two-replica 
partition function, summing over all independent secondary structures 
of a pair of RNAs of length $N-1$ bases in which there are exactly $n-1$
exterior bases of the common bond structure\footnote{An exterior base of
a secondary structure is a base that could be bound to a fictitious base
at position $N+1$ without disrespecting the no pseudo-knot constraint.}
all of which are completely unbound in both replicas. Since the
exterior bases form one of the bubbles of
the common bond structure, the possible binding configurations of
these exterior bases are described by $\tworQ(n)$. Thus, the full partition
function of the two replica problem can be calculated from this
restricted partition function as
\begin{equation}\label{eq_gbywandq}
\tworG(N;\widetilde{q})=\sum_{n=1}^N \tworW(N,n;\widetilde{q})\tworQ(n).
\end{equation}
Now, let us formulate a recursion relation for $\tworW$ by adding one 
additional base $N$ to each of the two RNAs.
We can separate the possible configurations of the new function
$\tworW(N+1,n;\widetilde{q})$ according to the possibilities that the new
base $N$ is either not involved in a common bond or forms a common
bond with base $1\le i<N$. This yields the recursion relation
\begin{eqnarray}\label{eq_trrecurs}
\lefteqn{\tworW(N+1,n;\widetilde{q})=}\hspace{5mm}\\\nonumber
&=&\tworW(N,n-1;\widetilde{q})+
q^2\widetilde{q}\sum_{i=n}^{N-1}\tworW(i,n;\widetilde{q})\tworG(N-i;\widetilde{q})
\end{eqnarray}
for $N\ge2$ and $n\ge1$. The applicable boundary
conditions are: $\tworW(N,N-1;\widetilde{q})=0$, $\tworW(N,N;\widetilde{q})=1$, and
$\tworW(N,n;\widetilde{q})\!=\!0$ for each $n\!>\!N\!\ge\!1$ and
$\tworW(N,0;\widetilde{q})\!=\!\delta_{N,0}$.

At this point, it is  convenient to introduce the $z$-transforms in
order to decouple the discrete convolution in
Eq.~(\ref{eq_trrecurs}). They are
\begin{eqnarray*}
\widehat{\tworG}(z;\widetilde{q})&\!\!\equiv\!\!&
\sum_{N=1}^\infty \tworG(N;\widetilde{q})z^{-N},\\
\widehat{\tworQ}(z)&\!\!\equiv\!\!&
\sum_{N=1}^\infty \tworQ(N)z^{-N},\quad\mbox{and}\\
\widehat{\tworW}(z,n;\widetilde{q})&\!\!\equiv\!\!&
\sum_{N=1}^\infty \tworW(N,n;\widetilde{q})z^{-N}\!\!=\!\!
\sum_{N=n}^\infty \tworW(N,n;\widetilde{q})z^{-N}\!\!\!.
\end{eqnarray*}
Using Eq.~(\ref{eq_trrecurs}) and the boundary conditions we get
\begin{eqnarray*}
\lefteqn{z\widehat{\tworW}(z,n;\widetilde{q})=}\hspace{3mm}\\
&=&\sum_{N=n}^\infty \tworW(N,n,\widetilde{q})z^{-(N-1)}\\
&=&z^{-{n-1}}+\sum_{N=n+1}^\infty \tworW(N+1,n;\widetilde{q})z^{-N}\\
&=&z^{-(n-1)}+\sum_{N=n+1}^\infty \tworW(N,n-1;\widetilde{q})z^{-N}\\
&&+q^2\widetilde{q}\!\!\!\sum_{N=n+1}^\infty\sum_{i=n}^\infty
\tworW(i,n;\widetilde{q})z^{-i}\tworG(N-i;\widetilde{q})z^{-(N-i)}\\
&=&\widehat{\tworW}(z,n-1;\widetilde{q})+
q^2\widetilde{q}\,\widehat{\tworW}(z,n;\widetilde{q})\widehat{\tworG}(z;\widetilde{q}).
\end{eqnarray*}
This can be solved for $\widehat{\tworW}(z,n;\widetilde{q})$ with the result
\begin{equation}
\widehat{\tworW}(z,n;\widetilde{q})=
\frac{1}{z-q^2\widetilde{q}\widehat{\tworG}(z;\widetilde{q})}
\widehat{\tworW}(z,n-1;\widetilde{q}).
\end{equation}
Together with the boundary condition $\widehat{\tworW}(z,0;\widetilde{q})=1$,
we get
\begin{equation}\label{eq_hatw}
\widehat{\tworW}(z,n;\widetilde{q})=
\big[z-q^2\widetilde{q}\widehat{\tworG}(z;\widetilde{q})\big]^{-n}.
\end{equation}
If we now multiply Eq.~(\ref{eq_gbywandq}) by $z^{-N}$ and sum both sides
over $N$ we get
\begin{equation}
\widehat{\tworG}(z;\widetilde{q})=
\sum_{n=1}^\infty\widehat{\tworW}(z,n;\widetilde{q})\tworQ(n)
\end{equation}
which, upon inserting Eq.~(\ref{eq_hatw}),  becomes an implicit equation
\begin{equation}\label{eq_trimplicit}
\widehat{\tworG}(z;\widetilde{q})=
\widehat{\tworQ}\big(z-q^2\widetilde{q}\widehat{\tworG}(z;\widetilde{q})\big)
\end{equation}
for the full partition function $\widehat{\tworG}(z;\widetilde{q})$, provided
that we know the function $\widehat{\tworQ}$.

Since $\widehat{\tworQ}$ does not depend on $\widetilde{q}$, we can find
its form using 
the following strategy: If $\widetilde{q}=1$, a common bond does not
contribute any additional Boltzmann factor. Thus, the two replica partition
function for this specific value of $\widetilde{q}$ is just the square
of the partition function of a single uniformly attracting RNA molecule,
i.e.,
\begin{equation}
\tworG(N;\widetilde{q}=1)=G^2(N). \label{eq_G2}
\end{equation}
Since we know $G(N)$ through the exact expression (\ref{eq_singlez}) 
for its $z$-transform $\widehat{G}$, we can regard
$\widehat{\tworG}(z;\widetilde{q}=1)$ 
as a known function, even though a closed form expression is not available.
From Eq.~(\ref{eq_trimplicit}), we have
\begin{equation}\label{eq_trimplicitqt1}
\widehat{\tworG}(z;\widetilde{q}=1)=
\widehat{\tworQ}\big(z-q^2\widehat{\tworG}(z;\widetilde{q}=1)\big).
\end{equation}
This is an equation for  $\widehat{\tworQ}$ in terms of the known 
function $\widehat{\tworG}(z;\widetilde{q}=1)$.
After we solve it for $\widehat{\tworQ}$ below, we can use
Eq.~(\ref{eq_trimplicit}) to solve for the only leftover unknown
$\widehat{\tworG}(z;\widetilde{q})$ for arbitrary values of
$\widetilde{q}$.

\subsection{Solution in the thermodynamic limit}

In the thermodynamic limit, it is sufficient to consider only the
singularities of the $z$-transform
$\widehat{\tworG}(z;\widetilde{q})$. From the form of
$\widehat{\tworG}$ in the vicinity of the singularity
$\zeta(\widetilde{q})$, the two-replica partition function
$\tworG(N;\widetilde{q})$ is readily obtained by the inverse
$z$-transform, with the result
\begin{equation}
\tworG(N;\widetilde{q}) = A(\widetilde{q}) N^{-\zexpo} \zeta^N(\widetilde{q}).
\end{equation}
The result immediately yields the free energy per length, 
$f = -k_BT \ln \zeta$. More significantly, the exponent $\zexpo$ reveals
which phase the two-replica system is in: for $\widetilde{q}=1$ (i.e.
no disorder), the two-replica system is just a product of 
two independent single-replica systems and we must have $\zexpo=3$
as implied by the single-replica partition function $G(N)$ in
Eq.~(\ref{eq_moltenz}).
On the other hand, for $\widetilde{q} \to \infty$, the two replicas are 
forced to be locked together and behave as a single replica. In this case,
 we must have  $\zexpo= 3/2$. As we will see, $\zexpo=3$ and $\zexpo=3/2$
are the only values this exponent can take on for this system; it indicates
whether or not   the two replicas are locked, and hence whether or not
the effect of disorder is relevant.

The singularity $\zeta(\widetilde{q})$ of
$\widehat{\tworG}(z;\widetilde{q})$ is given implicitly by
Eq.~(\ref{eq_trimplicit}), which we now analyze in detail.  We start
by recalling the solution of the homogeneous single RNA problem,
Eq.~(\ref{eq_moltenz}). From the relation (\ref{eq_G2}), we have 
$\tworG(N,\widetilde{q}=1)= A_0^2(q) N^{-2\zexpo_0} \singlezsing^{2N}$
for large $N$, with 
$\zexpo_0=3/2$, $\singlezsing=1+2\sqrt{q}$, and $A_0(q)$  given in
Sec.~\ref{sec_moltenpartfunc}.
Hence, the $z$-transform $\widehat{\tworG}(z;\widetilde{q}=1)
= \sum_N \tworG(N;\widetilde{q})z^{-N}$ is defined on the interval
$[\singlezsing^2,\infty[$. It is a monotonously decreasing function of $z$,
terminating with a singularity at $z=\singlezsing^2$ which produces the
$\zexpo=3$ singularity in $\tworG(N;\widetilde{q}=1)$.

Due to Eq.~(\ref{eq_trimplicitqt1}), the same singularity must occur in
$\widehat{\tworQ}(z)$ at $z=\Qsing\equiv \singlezsing^2-q^2 \gatsing$, where
\begin{equation}\label{eq_gatsingdef}
\gatsing\equiv\widehat{\tworG}(\singlezsing^2;\widetilde{q}=1)=
\sum_{N=1}^\infty G(N)^2\singlezsing^{-2N}\label{eq_defa}
\end{equation}
is a positive number and does not depend on anything else but $q$. Since
$z-q^2\widehat{\tworG}(z;\widetilde{q}=1)$ is a smooth monotonously increasing
function which maps the interval $[\singlezsing^2,\infty[$ into the interval
$[\Qsing,\infty[$, it follows from Eq.~(\ref{eq_trimplicitqt1})
that $\widehat{\tworQ}(z)$ is a smooth, monotonously decreasing function
which maps the interval $[\Qsing,\infty[$ into the interval $]0,\gatsing]$.

Now that we have characterized $\widehat{\tworQ}(z)$ in detail, we can
proceed to study $\widehat{\tworG}(z;\widetilde{q})$ for arbitrary
$\widetilde{q}$. Clearly, according to Eq.~(\ref{eq_trimplicit})
$\widehat{\tworG}(z;\widetilde{q})$ has a singularity leading to
$\zexpo=3$ at $z=\tworzsingmolten(\widetilde{q})$, defined
implicitly by
\begin{equation}
\tworzsingmolten(\widetilde{q})
-q^2\widetilde{q}\,\widehat{\tworG}(\tworzsingmolten(\widetilde{q});\widetilde{q})=\Qsing
\end{equation}
because $\widehat{\tworQ}(z)$ has this singularity at $z=\Qsing$.
Again according to Eq.~(\ref{eq_trimplicit}), we have
$\widehat{\tworG}(\tworzsingmolten(\widetilde{q});\widetilde{q})=\widehat{\tworQ}(\Qsing)=\gatsing$
independent of $\widetilde{q}$. This  leads to one of the key results
\begin{equation}\label{eq_moltenzresult}
\tworzsingmolten(\widetilde{q})=\Qsing+q^2\widetilde{q}\gatsing=
(1+2\sqrt{q})^2+q^2(\widetilde{q}-1)\gatsing.
\end{equation}

If $z=\tworzsingmolten(\widetilde{q})$ is the only singularity of 
$\widehat{\tworG}(z;\widetilde{q})$, it
would imply that there is only one phase with $\zexpo=3$,
and the free energy per length of the two-replica system is given by
\begin{equation}\label{eq_freeenergy0}
\tworfmolten=-k_BT\ln[(1+2\sqrt{q})^2+q^2(\widetilde{q}-1)\gatsing]
\end{equation}
for all values of $\widetilde{q}$.
By differentiating this with respect to $\widetilde{q}$, we obtain the
fraction of common contacts
\begin{equation}\label{eq_contacts0}
\tworccmolten=
\frac{q^2\widetilde{q}\gatsing}{(1+2\sqrt{q})^2+q^2\widetilde{q}\gatsing}
\end{equation}
in this phase
as a function of $\widetilde{q}$.  For very large disorder, i.e., for large
$\widetilde{q}$, this fraction converges to one. However, since it is
the fraction of bonds divided by the total number of bases and every
base pair has two bases, it has to be bounded from above by $1/2$. Thus, we
conclude that Eq.~(\ref{eq_freeenergy0}) cannot be the free energy of the
two-replica system for all $\widetilde{q}$'s. At least for large $\widetilde{q}$,
there must be
another singularity of $\widehat{\tworG}(z;\widetilde{q})$ which will yield a 
different expression for the free energy to give
physically reasonable fraction of common bonds.

In order to find this other singularity, we introduce the inverse function
$\widehat{\tworZ}(g;\widetilde{q})$ of %the Laplace transformed partition function 
$\widehat{\tworG}(z;\widetilde{q})$. From Eq.~(\ref{eq_trimplicit}), it
follows that for any $\widetilde{q}$ and any $g\in]0,\gatsing]$,
\begin{eqnarray*}
\widehat{\tworG}(\widehat{\tworZ}(g;\widetilde{q});\widetilde{q})&=&
\widehat{\tworQ}[\widehat{\tworZ}(g;\widetilde{q})-q^2\widetilde{q}
\widehat{\tworG}(\widehat{\tworZ}(g;\widetilde{q});\widetilde{q})]\\
\Rightarrow\quad g&=&
\widehat{\tworQ}[\widehat{\tworZ}(g;\widetilde{q})-q^2\widetilde{q} g]\\
\Rightarrow\quad \widehat{\tworQ}^{-1}(g)&=&
\widehat{\tworZ}(g;\widetilde{q})-q^2\widetilde{q} g.
\end{eqnarray*}
Since $\widehat{\tworQ}^{-1}(g)$ does not depend on $\widetilde{q}$, we can 
eliminate it by evaluating the last equation above at the special value
$\widetilde{q}=1$ and write
\begin{equation}\label{eq_zrelation}
\widehat{\tworZ}(g;\widetilde{q})=\widehat{\tworZ}(g;1)+q^2(\widetilde{q}-1)g,
\end{equation}
where $\widehat{\tworZ}(g;1)$ is the inverse of the known function
 $\widehat{\tworG}(z;\widetilde{q}=1)$.
Eq.~(\ref{eq_zrelation}) is now an explicit solution 
for the inverse of $\widehat{\tworG}(z;\widetilde{q})$ for
{\em arbitrary} $\widetilde{q}$, and the singularity of $\widehat{\tworZ}$, 
located at $g=\tworgatsingglass(\widetilde{q})$ and
$\widehat{\tworZ}(\tworgatsingglass,\widetilde{q})=\tworzsingglass(\widetilde{q})$,
yields the free energy of the two-replica system, i.e.,
$\tworfglass= -k_BT \ln \tworzsingglass$, in this phase.

\begin{figure}[htbp]
\begin{center}
\includegraphics[angle=0,width=1.0\columnwidth]{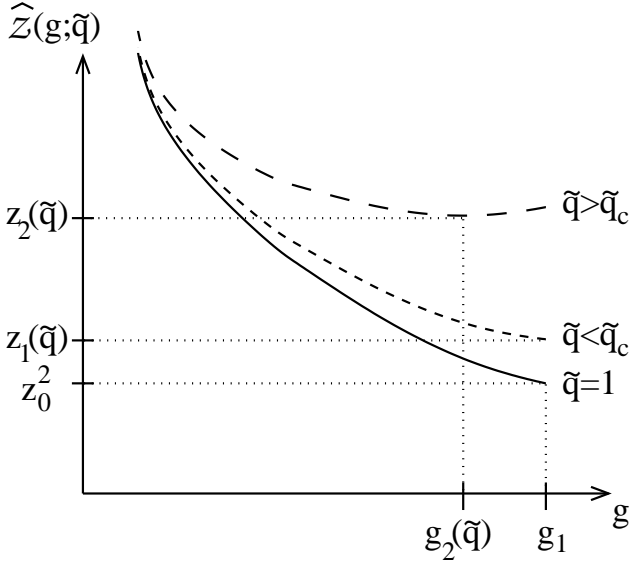}
\caption{Inverse of the Laplace transformed partition function
\protect$\widehat{\tworZ}(g;\widetilde{q})$ of the two replica system at
various values of the common bond interaction
\protect$\widetilde{q}$: The solid line shows the free system without
any interaction of common bonds.  In the presence of an interaction,
the inverse function of the partition function can be obtained by
adding a linear function to the free system function. If the
interaction is not too strong
(\protect$\widetilde{q}<\widetilde{q}_{\mathrm{c}}$, short dashed
line) adding the linear function with a small slope does not change
the qualitative form of the partition function. In this case, the two
replica system is controlled by the singularity at $g=\gatsing$ which is
independent of \protect$\widetilde{q}$. At stronger interactions
(\protect$\widetilde{q}>\widetilde{q}_{\mathrm{c}}$, long dashed line)
the inverse function develops a minimum. Beyond this minimum it is not
invertible any more and the two replica system is then dominated by
the singularity arising from this minimum.}
\label{fig_zhatsketch}
\end{center}
\end{figure}

While App.~\ref{app_derivez1} derives the position of the dominant
singularity present in Eq.~(\ref{eq_zrelation}) rigorously, we will
resort to some intuitive argument here. Since $\widehat{\tworG}(z;1)$
is a monotonously decreasing, convex function, so is its inverse
$\widehat{\tworZ}(g;1)$.  The latter function ends at the point
$(\singlezsing^2,\gatsing)$ with some slope $\slope<0$ in a
singularity which produces the $\zexpo=3$ behavior indicative of two
independently fluctuating uniformly self-attracting replicas. This is
shown in Fig.~\ref{fig_zhatsketch} as the solid line. According to
Eq.~(\ref{eq_zrelation}), we can obtain the corresponding function
$\widehat{\tworZ}(g;\widetilde{q})$ for arbitrary $\widetilde{q}$ by
simply adding a linear function $q^2(\widetilde{q}-1)g$ to this
function. If the slope $q^2(\widetilde{q}-1)$ of this linear function
is less than the smallest slope of $\widehat{\tworZ}(g;1)$, i.e., if
$\slope+q^2(\widetilde{q}-1)<0$, adding this linear function does not
qualitatively change anything (see the short dashed line in
Fig.~\ref{fig_zhatsketch}.)  The only singularity is still the one at
$g=\gatsing$, the corresponding value in $z$, i.e.,
$\tworzsingmolten(\widetilde{q})=\widehat{\tworZ}(\gatsing;\widetilde{q})$,
is trivially shifted by an amount $q^2(\widetilde{q}-1)\gatsing$ from
$\singlezsing^2$ as $\widetilde{q}$ varies. Thus, the scaling behavior
is characterized by $\zexpo=3$ as if $\widetilde{q}=1$ (i.e., absence
of disorders), although the free energy
$\tworfmolten=-k_BT\ln\tworzsingmolten(\widetilde{q})$ is shifted as already
derived in Eq.~(\ref{eq_freeenergy0}).

If the final slope $\slope+q^2(\widetilde{q}-1)$ of the right hand
side of Eq.~(\ref{eq_zrelation}) is {\em positive} the situation
becomes much different: Upon adding the linear function,
$\widehat{\tworZ}(g;\widetilde{q})$ develops a minimum at some
position
$(\tworgatsingglass(\widetilde{q}),\tworzsingglass(\widetilde{q}))$. Thus,
the inverse function $\widehat{\tworG}(z;\widetilde{q})$ has to be
calculated from the left (small $g$) branch of
$\widehat{\tworZ}(g;\widetilde{q})$ and has a {\em square root
singularity} at $\tworzsingglass(\widetilde{q})$. This square root
singularity implies that the characteristic exponent becomes
$\zexpo=3/2$, consistent with the picture that in this phase the two
replicas are locked together and fluctuate as one single effective RNA
molecule.

The position $\tworgatsingglass(\widetilde{q})$ of the minimum is
determined by the root of the derivative, i.e., by
\begin{equation}
-q^2(\widetilde{q}-1)=
\left.\frac{\mathrm{d}}{\mathrm{d}g}\right|_{g=\tworgatsingglass(\widetilde{q})}
\!\!\!\!\!\!\!\!\widehat{\tworZ}(g;1)=
\left[\left.\frac{\mathrm{d}}{\mathrm{d}g}
\right|_{z=\widehat{\tworZ}(\tworgatsingglass(\widetilde{q});1)}\!\!\!\!\!\!\!\!\!\!\!
\widehat{\tworG}(z;1)\right]^{-1}\!\!\!\!\!\!\!\!.
\end{equation}
The corresponding value
$\widehat{\tworZ}(\tworgatsingglass(\widetilde{q});\widetilde{q})$
determines the location of the square root singularity
$\tworzsingglass(\widetilde{q})$ of
$\widehat{\tworG}(z;\widetilde{q})$, i.e., the free energy per
length of the two replica problem.

$\tworzsingglass(\widetilde{q})$ can be conveniently expressed in
terms of the auxiliary quantity $z_{\mathrm{c}}(\widetilde{q})$
defined through
$\tworgatsingglass(\widetilde{q})=\widehat{\tworG}(z_{\mathrm{c}}(\widetilde{q});1)$
as
\begin{eqnarray*}
\tworzsingglass(\widetilde{q})&=&
\widehat{\tworZ}(\tworgatsingglass(\widetilde{q});\widetilde{q})\\
&=&\widehat{\tworZ}(\tworgatsingglass(\widetilde{q});1)
+q^2(\widetilde{q}-1)\tworgatsingglass(\widetilde{q})\\
&=&z_{\mathrm{c}}(\widetilde{q})
+q^2(\widetilde{q}-1)\widehat{\tworG}(z_{\mathrm{c}}(\widetilde{q});1).
\end{eqnarray*}
Comparing this expression with Eq.~(\ref{eq_freeenergy0}) which
is valid for small $\widetilde{q}$, we can
summarize the complete solution in terms of
\begin{equation}\label{eq_zkdef}
z_{\mathrm{c}}(\widetilde{q})\!=\!\!\left\{\begin{array}{ll}\mbox{the unique
$z\in]\singlezsing^2,\infty[$ that ful-}\ 
&\widetilde{q}>\widetilde{q}_{\mathrm{c}}\\
\mbox{fils }\frac{\mathrm{d}}{\mathrm{d}z}\widehat{\tworG}(z;1)
\!=\!\!-\!1\!/[q^2(\widetilde{q}\!-\!1)]&\\[10pt]
\singlezsing^2=(1+2\sqrt{q})^2&\widetilde{q}\le\widetilde{q}_{\mathrm{c}}
\end{array}\right.
\end{equation}
where
\begin{equation}\label{eq_qcdef}
\widetilde{q}_{\mathrm{c}}\equiv1-\frac{\slope}{q^2}>1,
\end{equation}
and 
\begin{equation}
\slope\equiv
\frac{1}{\left.\frac{\mathrm{d}}{\mathrm{d}z}\right|_{z=\singlezsing^2}
\widehat{\tworG}(z;\widetilde{q}=1)}
=-\frac{1}{\sum_{N=1}^\infty N G(N)^2\singlezsing^{-2(N-1)}}.
\end{equation}

In terms of this $z_c(\widetilde{q})$, the smallest singularity of
$\widehat{\tworZ}(z;\widetilde{q})$ is located at
\begin{equation}\label{eq_zetaresult}
\zeta(\widetilde{q})=z_{\mathrm{c}}(\widetilde{q})+q^2(\widetilde{q}-1)
\widehat{\tworG}(z_{\mathrm{c}}(\widetilde{q});1).
\end{equation}
The free energy per length of the two replica system is given by
\begin{equation}\label{eq_tworepfreeenergy}
f=-k_BT\ln[z_{\mathrm{c}}(\widetilde{q})+q^2(\widetilde{q}-1)
\widehat{\tworG}(z_{\mathrm{c}}(\widetilde{q});1)]
\end{equation}
and the fraction of common bonds is
\begin{equation}
s=\frac{q^2\widetilde{q}\widehat{\tworG}(z_{\mathrm{c}}(\widetilde{q});1)}%
{z_{\mathrm{c}}(\widetilde{q})+q^2(\widetilde{q}-1)
\widehat{\tworG}(z_{\mathrm{c}}(\widetilde{q});1)}.
\end{equation}
This fraction of common bonds turns out to be continuous at the phase
transition but it exhibits a jump in its slope at
$\widetilde{q}=\widetilde{q}_c$.

In the case $\widetilde{q}\le\widetilde{q}_{\mathrm{c}}$, these
simplify to Eqs.~(\ref{eq_freeenergy0}) and~(\ref{eq_contacts0}) (or
Eqs.~(\ref{eq_moltenzinmaintext}) and~(\ref{eq_defginmaintext})
respectively), with $z_{\mathrm{c}}(\widetilde{q})=\singlezsing^2$
independent of $\widetilde{q}$. The type of singularity of the
Laplace transformed partition function is the same as at
$\widetilde{q}=1$, resulting in $\zexpo=3$.
For $\widetilde{q}>\widetilde{q}_{\mathrm{c}}$, we cannot write down a
closed form expression for $\zeta(\widetilde{q})$ any more. But it is
given implicitly in terms of the solution of Eq.~(\ref{eq_zkdef}); it
involves only single replica quantities and can thus be evaluated
numerically.  Moreover, we have seen that the dominant singularity is
in this regime a square root singularity implying $\zexpo=3/2$.

\section{The free energy of the two-replica
problem}\label{app_derivez1}

In this appendix we give a derivation of the position of the
non-trivial singularity in the Laplace transform of the partition
function $\widehat{\tworG}(z;\widetilde{q})$. A more intuitive, graphical
derivation of this result was given in App.~\ref{app_solution}. Using
Eq.~(\ref{eq_zrelation}), we start by calculating
\begin{eqnarray*}
\frac{\mathrm{d}}{\mathrm{d}z}\widehat{\tworG}(z;\widetilde{q})&=&
\left[\left.\frac{\mathrm{d}}{\mathrm{d}g}
\right|_{g=\widehat{\tworG}(z;\widetilde{q})}\!\!\!\!\widehat{\tworZ}(g;\widetilde{q})
\right]^{-1}\\
&=&\left[\left.\frac{\mathrm{d}}{\mathrm{d}g}
\right|_{g=\widehat{\tworG}(z;\widetilde{q})}\!\!\!\!
\widehat{\tworZ}(g;1)+q^2(\widetilde{q}-1)
\right]^{-1}\\
&=&\left[\left[\left.\frac{\mathrm{d}}{\mathrm{d}z}
\right|_{z=\widehat{\tworZ}(\widehat{\tworG}(z;\widetilde{q});1)}\!\!\!\!\widehat{\tworG}(z;1)
\right]^{-1}\!\!\!\!\!\!\!\!\!+q^2(\widetilde{q}-1)\right]^{-1}\!\!\!\!.
\end{eqnarray*}
This expression obviously has a singularity at
$\tworzsingglass(\widetilde{q})$ which is defined by
\begin{equation}\label{eq_z1def}
\left.\frac{\mathrm{d}}{\mathrm{d}z}
\right|_{z=\widehat{\tworZ}(\widehat{\tworG}(\tworzsingglass(\widetilde{q});
\widetilde{q});1)}
\widehat{\tworG}(z;1)=-\frac{1}{q^2(\widetilde{q}-1)}.
\end{equation}
Since
$\frac{\mathrm{d}}{\mathrm{d}z}\widehat{\tworG}(z;1)\in[1/\slope,0[$,
this is only possible for
\begin{displaymath}
\widetilde{q}\ge\widetilde{q}_{\mathrm{c}}\equiv1-\frac{\slope}{q^2}.
\end{displaymath}
For smaller values of $\widetilde{q}$, there is no other singularity and
the free energy per length is given by Eq.~(\ref{eq_freeenergy0}).

If $\widetilde{q}\ge\widetilde{q}_{\mathrm{c}}$ the additional
singularity $\tworzsingglass(\widetilde{q})$ exists and is --- as we
will see below --- always smaller than the singularity
$\tworzsingmolten(\widetilde{q})$. Thus, the free energy per length is
given by the singularity $\tworzsingglass(\widetilde{q})$ in the
strong coupling phase, i.e., for
$\widetilde{q}\ge\widetilde{q}_{\mathrm{c}}$.

On the first sight, Eq.~(\ref{eq_z1def}) still looks as if
$\tworzsingglass(\widetilde{q})$ could only be calculated if the full function
$\widehat{\tworG}(z;\widetilde{q})$ is known. However, for any
$\widetilde{q}\ge\widetilde{q}_{\mathrm{c}}$ we can define
$z_{\mathrm{c}}(\widetilde{q})$ to be the unique solution of the equation
\begin{displaymath}
\left.\frac{\mathrm{d}}{\mathrm{d}z}\right|_{z=z_{\mathrm{c}}(\widetilde{q})}
\widehat{\tworG}(z;1)=-\frac{1}{q^2(\widetilde{q}-1)}.
\end{displaymath}
This quantity depends only on the  function
$\widehat{\tworG}(z;1)$. According to Eq.~(\ref{eq_z1def}), $\tworzsingglass(\widetilde{q})$
and $z_{\mathrm{c}}(\widetilde{q})$ are related by
$\widehat{\tworZ}(\widehat{\tworG}(\tworzsingglass(\widetilde{q});\widetilde{q});1)=
z_{\mathrm{c}}(\widetilde q)$. This implies that 
$\widehat{\tworG}(\tworzsingglass(\widetilde{q});\widetilde{q})=
\widehat{\tworG}(z_{\mathrm{c}}(\widetilde{q});1)$. On the other hand,
Eq.~(\ref{eq_zrelation}) applied to
$g=\widehat{\tworG}(\tworzsingglass(\widetilde{q});\widetilde{q})$ yields
\begin{eqnarray*}
\tworzsingglass(\widetilde{q})&=&
\widehat{\tworZ}(\widehat{\tworG}(\tworzsingglass(\widetilde{q});\widetilde{q});\widetilde{q})\\
&=&\widehat{\tworZ}(\widehat{\tworG}(\tworzsingglass(\widetilde{q});\widetilde{q});1)
+q^2(\widetilde{q}-1)\widehat{\tworG}(\tworzsingglass(\widetilde{q});\widetilde{q})\\
&=&z_{\mathrm{c}}(\widetilde{q})+q^2(\widetilde{q}-1)
\widehat{\tworG}(z_{\mathrm{c}}(\widetilde{q});1)
\end{eqnarray*}
which is finally an expression which involves only quantities of the
non-interacting system. Since
$z+q^2(\widetilde{q}-1)\widehat{\tworG}(z;1)=
\widehat{\tworZ}(\widehat{\tworG}(z;1);\widetilde{q})$ is a monotonous
function on the interval
$[\singlezsing^2,z_{\mathrm{c}}(\widetilde{q})]$, we always have
$\tworzsingglass(\widetilde{q})\le \tworzsingmolten(\widetilde{q})$ with
equality if and only if
$z_{\mathrm{c}}(\widetilde{q})=\singlezsing^2$, i.e., for
$\widetilde{q}=\widetilde{q}_{\mathrm{c}}$. Therefore, the free energy
per length is indeed given by
\begin{displaymath}
\tworfglass=-k_BT\ln[z_{\mathrm{c}}(\widetilde{q})+q^2(\widetilde{q}-1)
\widehat{\tworG}(z_{\mathrm{c}}(\widetilde{q});1)]
\end{displaymath}
for any $\widetilde{q}\ge\widetilde{q}_{\mathrm{c}}$.

\end{document}